\documentclass[twocolumn,aps,showpacs,prb]{revtex4-2}

\usepackage{graphicx}
\usepackage{epstopdf}
\usepackage{amsmath}
\usepackage{multirow}
\usepackage{xcolor}

\begin{document}

\title{Microscopic resolution of superconducting electrons in ultrahigh-pressed hydrogen sulfide}

\author{Jian-Feng Zhang$^1$}
\author{Bo Zhan$^1$}
\author{Miao Gao$^2$}
\author{Kai Liu$^3$}
\author{Xinguo Ren$^1$}\email{renxg@iphy.ac.cn}
\author{Zhong-Yi Lu$^3$}\email{zlu@ruc.edu.cn}
\author{Tao Xiang$^{1,4,5}$}\email{txiang@iphy.ac.cn}

\affiliation{$^1$Institute of Physics, Chinese Academy of Sciences, Beijing 100190, China}
\affiliation{$^2$Department of Physics, School of Physical Science and Technology, Ningbo University, Zhejiang 315211, China}
\affiliation{$^3$Department of Physics and Beijing Key Laboratory of Opto-electronic Functional Materials $\&$ Micro-nano Devices, Renmin University of China, Beijing 100872, China}
\affiliation{$^4$Beijing Academy of Quantum Information Sciences, Beijing 100193, China}
\affiliation{$^5$School of Physical Sciences, University of Chinese Academy of Sciences, Beijing 100049, China}

\date{\today}

\begin{abstract}
  We investigate the electronic and phonon properties of hydrogen sulfide (SH$_3$) under ultrahigh pressure to elucidate the origin of its high-T$_c$ superconductivity. Contrary to the prevailing belief that the metalized S-H $\sigma$ bond is responsible, our analysis, based on the anisotropic Migdal-Eliashberg equation and the crystal orbital Hamilton population (COHP) calculation, reveals that the H-H $\sigma$-antibonding states play a dominant role in the large electron-phonon coupling that leads to the superconducting pairing in SH$_3$. Furthermore, by partially restricting the vibration of S atoms, we demonstrate that the S-H bonds provide subsidiary contributions to the pairing interaction. These findings shed light on the importance of the previously overlooked H-H $\sigma^*$ bonds in driving high-T$_c$ superconductivity in SH$_3$ and offer insights into the relationship between metallic H-H covalent antibonding and high-T$_c$ superconductivity in other hydrogen-rich materials under high pressure.
\end{abstract}

\pacs{}

\maketitle

\section{INTRODUCTION}

 The search for high-temperature superconductors is a critical objective in superconductivity research, as it holds significant theoretical and practical implications. While conventional superconductivity is well-understood in terms of electron-phonon coupling (EPC) within the framework of the Bardeen-Cooper-Schrieffer (BCS) theory, unconventional superconductivity, found in cuprates~\cite{cup1, cup2, cup3, cup4}, iron-based superconductors~\cite{iro1,iro2,iro3,iro4,iro5,iro6}, and heavy-fermion compounds~\cite{hea1,hea2,hea3,hea4}, presents a more complex challenge that is believed to be intimately connected to spin fluctuations~\cite{iro2,spinflu}. In the BCS theory~\cite{bcs}, the superconducting transition temperature (T$_c$) is primarily determined by three factors: the electronic density of states (DOS) at the Fermi level, the EPC strength, and the phonon frequency with strong EPC. However, unconventional superconductors have much higher transition temperatures than those predicted by BCS theory, suggesting the presence of other factors that are not yet fully understood. One promising avenue of research is exploring the use of materials containing light elements, such as hydrogen, which are expected to possess higher superconducting T$_c$ values due to their ability to generate high-frequency phonons. Metallic hydrogen, in particular, has been predicted to have an ultra-high T$_c$~\cite{metalHpre, metalHcal}, but this has not yet been achieved due to the limitations of high-pressure technology~\cite{presstec}.

 Furthermore, metalizing $\sigma$-bonding or other strong chemical bonding electrons also provides a universal approach to forming high-T$_c$ pairing~\cite{gaoprb2015,gao2015}. A $\sigma$-covalent bond involves a spin singlet pair of electrons, and the attractive force between these paired electrons is typically strong. Consequently, a $\sigma$ bond forms a stable bound state below the Fermi level in most materials, not contributing to electrical conductivity. However, it is feasible to attenuate the binding force and metalize $\sigma$-bonding electrons by applying pressure, doping, crystal field effects, or other approaches. This metalization process unavoidably weakens the coupling between the $\sigma$-bonding electrons. Nevertheless, even with this reduced coupling, the remaining interactions among these electrons may still be sufficiently strong to support the formation of high-T$_c$ superconductivity.

 In line with this scenario, MgB$_2$ exhibits the highest superconducting T$_c$ among all conventional BCS superconductors discovered at ambient pressure, with a value of 39 K~\cite{MgB2_1st}. Hydrogen-rich materials, SH$_3$\cite{SH3_1st}, LaH$_{10}$\cite{LaH10_1st,LaH10_2nd}, YH$_9$\cite{YH9_1st}, and CaH$_6$\cite{CaH6_1st, CaH6_2nd}, were reported to exhibit high-T$_c$ superconductivity above 100 K under ultrahigh pressure. These high-T$_c$ hydrides were also found to have metalized $\sigma$-covalent electrons involving hydrogen atoms~\cite{maprl, maprl2, MH_review2, SH3_pre, CaH6_pre}.

 Compressed SH$_3$ was the first hydride to exhibit high-T$_c$ superconductivity, with a temperature as high as 203 K at pressures exceeding 150 GPa \cite{SH3_1st}. At pressures above 150 GPa, SH$_3$ adopts a body-centered cubic (bcc) crystal structure with a space group of \textit{Im-3m} (No. 229) \cite{SH3str1, SH3str2, SH3str3}. Both isotope effect experiments and DFT calculations suggest that SH$_3$ is a conventional BCS-type superconductor \cite{SH3_1st, SH3_pre}. In MgB$_2$, both the B-B $\sigma$ and $\pi$ bonding bands around the Fermi level contribute to the superconductivity \cite{mgb2-pcs, mgb2-stm, mgb2-arpes, mgb2-calnature}. In SH$_3$, the metalized S-H covalent bond with strong electron-phonon coupling (EPC) was found to have a substantial contribution to the pairing interaction \cite{SH3str1, SH3_pre, SH3_1st}, similar to MgB$_2$ where strong EPC arises from the metalized B-B $\sigma$ bonds that strongly couple with B stretching vibration modes \cite{MgB2_sigma}. However, the effect of metalized H-H covalent bonds on the pairing interaction in SH$_3$ is unknown.

 This paper investigates the chemical bonding properties between neighboring atoms in SH$_3$ at 200 GPa through non-orthogonal atomic orbital-based electronic structure calculations. Our analysis, using the crystal orbital overlap population (COOP) and the crystal orbital Hamilton population (COHP) techniques~\cite{coop, coop2}, reveals that the electronic states around the Fermi level exhibit both S-H and H-H covalent bonding/antibonding states. We further solve the anisotropic Migdal-Eliashberg equation~\cite{ani-ME} to determine the EPC strength $\lambda_{n{\bf k}}$ and the superconducting energy gap $\Delta_{n{\bf k}}$ of electrons on the Fermi surfaces. By comparing the bonding populations and anisotropic superconducting gaps around the Fermi level, we find that the electronic states with larger $\lambda_{n{\bf k}}$ and $\Delta_{n{\bf k}}$ are primarily composed of H-H covalent antibonding states, with a small contribution from other bonding states. Our findings challenge the conventional picture that high-T$_c$ superconductivity in SH$_3$ is primarily driven by metalized S-H covalent bonds~\cite{SH3_pre, SH3str1, SH3_1st}, instead suggesting that the H-H covalent antibonds play a more decisive role in the superconducting pairing in this material.

\section{Calculation Methods}

 The electronic structure and chemical bonding properties of SH$_3$ at 200 GPa were investigated using density functional theory (DFT)~\cite{dft1,dft2} and the Fritz Haber Institute \textit{ab initio molecular simulations} (FHI-aims) package~\cite{fhi-aims}. FHI-aims is an all-electron, full-potential electronic structure code that uses non-orthogonal atom-centered basis functions. The Perdew-Burke-Ernzerhof (PBE) generalized gradient approximation (GGA)~\cite{PBE} was used to evaluate the exchange-correlation energy. To sample the Brillouin zone, a 24$\times$24$\times$24 {\bf k}-point mesh was used. The lattice constants were converged up to a numerical precision of 0.01 \AA, and total energies with a precision of less than 1 meV/atom. The atomic forces were converged to an accuracy of 10$^{-4}$ eV/\AA\ and the stress to 10$^{-3}$ kbar. After geometry relaxation at 200 GPa, the SH$_3$ crystal showed a perfect bcc lattice structure with the shortest H-H and S-H bond lengths of 1.49 \AA.

 In FHI-aims, the eigenfunctions are represented by $\psi_{n{\bf k}}(\textbf r)$ and expanded using numeric atom-centered orbitals (NAOs). These NAOs are grouped into hierarchical sets, known as ``tiers", which range from a minimal basis set to larger ones that achieve meV/atom accuracy in total energies. For the purpose of this work, we utilized the ``tier 1" basis set. Specifically, for H, we used two radial functions for the $s$ channel and one for the $p$ channel, while for S, we used two radial functions for both $s$ and $p$ channels and one for the $d$ and $f$ channels to describe the valence states. The primary minimal basis set is formed by one atomic-like $s$-type function for H and one $s$- and one $p$-type radial functions for both H and S, with additional hydrogen- and ionic-type radial functions \cite{fhi-aims} included to provide extra variational flexibility and enhance accuracy. To facilitate chemical bonding analysis, we expressed the Kohn-Sham wave functions as
\begin{equation}
|n{\bf k}\rangle =\sum_{i}C_{i}^{n\bf k}|{\bf k}i\rangle+|{\bf A}^{n{\bf k}}\rangle ,
\label{eq:KS_expansion}
\end{equation}
 where the first term includes contributions from the minimal atomic basis set mentioned earlier, and the second term, $|{\bf A}^{n{\bf k}}\rangle$, includes contributions from additional basis functions. $C_{i}^{n\bf k}$ is the expansion coefficient of the atomic orbital basis in $\bf k$ space which defined by
\begin{equation}
|{\bf k}i\rangle =\sum_{R}{\rm {exp}}({\rm i}{\bf k\cdot\textbf R})|{\bf R}i\rangle
\label{eq:F_expansion}
\end{equation}
 and $|{\bf R}i\rangle$ represents an atomic orbital within a unit cell defined by the lattice vector ${\bf R}$. To simplify the analysis, only the minimal atomic orbitals $|{\bf R}i\rangle$ are taken into account in the bonding population analysis.

 We analyzed the chemical bonding characteristics of electronic states surrounding the Fermi level. Specifically, the analysis focuses on the hybridization between local atomic orbitals $|{\bf R}i\rangle$ and $|{\bf R'}i'\rangle$ within the Bloch state $|n{\bf k}\rangle$. This hybridization can be quantified by two quantities, COOP and COHP~\cite{coop,coop2}, defined as follows:
\begin{eqnarray}
{\text {COOP}}^{n\bf k}_{{\bf R'}i' {\bf R}i} &= &
 {C_{i'}^{n\bf k}}^*C_{i}^{n\bf k} e^{ i \mathbf{k} \cdot (\mathbf{R} - \mathbf{R}' )} \langle \mathbf{R}' i' | \mathbf{R} i\rangle, \\
{\text {COHP}}^{n\bf k}_{{\bf R'}i' {\bf R}i} &= &
-{C_{i'}^{n\bf k}}^*C_{i}^{n\bf k} e^{ i \mathbf{k} \cdot (\mathbf{R} - \mathbf{R}' )} \langle{\bf R'}i'|\hat{H}|{\bf R} i\rangle .
\label{eq_coopcohp}
\end{eqnarray}
 The real parts of COOP and COHP enable us to differentiate between bonding and antibonding states, while their magnitudes reflect the importance of a given pair of atomic orbitals in the chemical bonding of an electronic state. In general, we focused on its summation in energy space:
\begin{eqnarray}
{\text {COOP}}_{{\bf R'}i' {\bf R}i}(\varepsilon) &= &
\sum_{n\bf k}{{\text {COOP}}^{n\bf k}_{{\bf R'}i' {\bf R}i}\delta(\varepsilon-E_{n\bf k})}, \\
{\text {COHP}}_{{\bf R'}i' {\bf R}i}(\varepsilon) &= &
\sum_{n\bf k}{{\text {COHP}}^{n\bf k}_{{\bf R'}i' {\bf R}i}\delta(\varepsilon-E_{n\bf k})}.
\label{eq_coopcohp}
\end{eqnarray}

 We utilized the density functional perturbation theory (DFPT)\cite{dfptreview, dfptreview2} implemented in the Quantum ESPRESSO (QE) package\cite{pwscf} to compute the phonon and EPC properties of SH$_3$. The interactions between the electrons and nuclei were described using norm-conserving pseudopotentials~\cite{ncpp}, while the GGA approximation of the PBE type~\cite{PBE} was used for the exchange-correlation functional. We set the plane-wave basis kinetic energy cutoff to ensure accuracy to 80 Ry. For electronic and phonon structure calculations, we sampled BZ using a 24$\times$24$\times$24 {\bf k}-point mesh and an 8$\times$8$\times$8 {\bf q}-point mesh, respectively. Finally, we employed the Gaussian smearing method with a 0.004 Ry width for Fermi surface broadening.

 We employed the Migdal-Eliashberg theory to investigate the superconducting properties of SH$_3$. Specifically, we utilized the EPW package~\cite{epw}, which is renowned for its ability to accurately capture electronic, phononic, and EPC properties by implementing maximally localized Wannier functions (MLWF)~\cite{mlwf}. Our calculations involved the use of an 8$\times$8$\times$8 momentum mesh, which was subsequently interpolated to a denser mesh of 32$\times$32$\times$32 to enhance the accuracy of our results. By implementing this meticulous approach, we were able to obtain a detailed understanding of the superconducting properties of SH$_3$, including its anisotropic behavior on the Fermi surfaces, by self-consistently solving the anisotropic Migdal-Eliashberg equation~\cite{ani-ME}:
\begin{eqnarray}
 Z\left(n\mathbf{k}, i \omega_{m}\right)
&= & 1+ \frac{\pi T}{N(\text{E}_\text{F})\omega_{m}} \sum_{n'\mathbf{k}^{\prime} m^{\prime}} \delta(E_{{n\bf k}'}-E_\text{F}) \nonumber \\
&& \times \frac{\omega_{m^{\prime}} \lambda({n\mathbf{k}, n'\mathbf{k}^{\prime}},\omega_m-\omega_{m'}) }{\sqrt{\omega_{m'}^{2}+\Delta^{2}\left(n'\mathbf{k}', i \omega_{m'}\right)}} ,
\end{eqnarray}
and
\begin{eqnarray}
&& Z\left(n\mathbf{k}, i \omega_{m}\right) \Delta\left(n\mathbf{k}, i \omega_{m}\right) \nonumber \\
&=& \frac{\pi T}{N(E_\text{F})} \sum_{n'\mathbf{k}^{\prime} m^{\prime}} \frac{\Delta\left(n'\mathbf{k}^{\prime}, i \omega_{m^{\prime}}\right) \delta(E_{{n'\bf k}'}-E_\text{F}) }{\sqrt{\omega_{m'}^{2}+\Delta^{2}\left(n'\mathbf{k}', i \omega_{m'}\right)}} \nonumber \\
&& \times[\lambda({n\mathbf{k}, n'\mathbf{k}^{\prime}},\omega_m-\omega_{m'})-N(E_\text{F})V\left(\mathbf{k}-\mathbf{k}^{\prime}\right)] ,
\end{eqnarray}
where $\lambda({n\mathbf{k}, n'\mathbf{k}^{\prime}},\omega_m-\omega_{m'})$ is an auxiliary function that describes the anisotropic electron-phonon coupling:
\begin{equation}
 \lambda({n\mathbf{k}, n'\mathbf{k}^{\prime}},\omega_m-\omega_{m'})
 = \sum_\nu\frac{2\omega_{\bf k-\bf k'\nu} |g_{{n\bf k}n'\bf k'\nu}|^2 }{(\omega_m-\omega_{m'})^2 + \omega_{\bf k-\bf k'\nu}^2}  .
\label{eq:aux_EPC}
\end{equation}
Here, $N(E_\text{F})$ is the total DOS at the Fermi level, $Z\left(n\mathbf{k}, i \omega_{m}\right)$ is the renormalization factor and $\Delta\left(n\mathbf{k}, i \omega_{m}\right)$ is the superconducting gap, $\omega_{m}=(2m+1)\pi T$ is the Fermion Matsubara frequency, and $V\left(\mathbf{k}-\mathbf{k}^{\prime}\right)$ is a screened Coulomb interaction. As an approximation, $N(E_\text{F})V\left(\mathbf{k}-\mathbf{k}^{\prime}\right)$ is replaced by a semiempirical constant $\mu^*$=0.16, following Ref.~[\onlinecite{ani-ME}].

\section{Results and Discussion}

\subsection{Crystal and electronic structure}

\begin{figure}[tb]
\includegraphics[angle=0,scale=0.43]{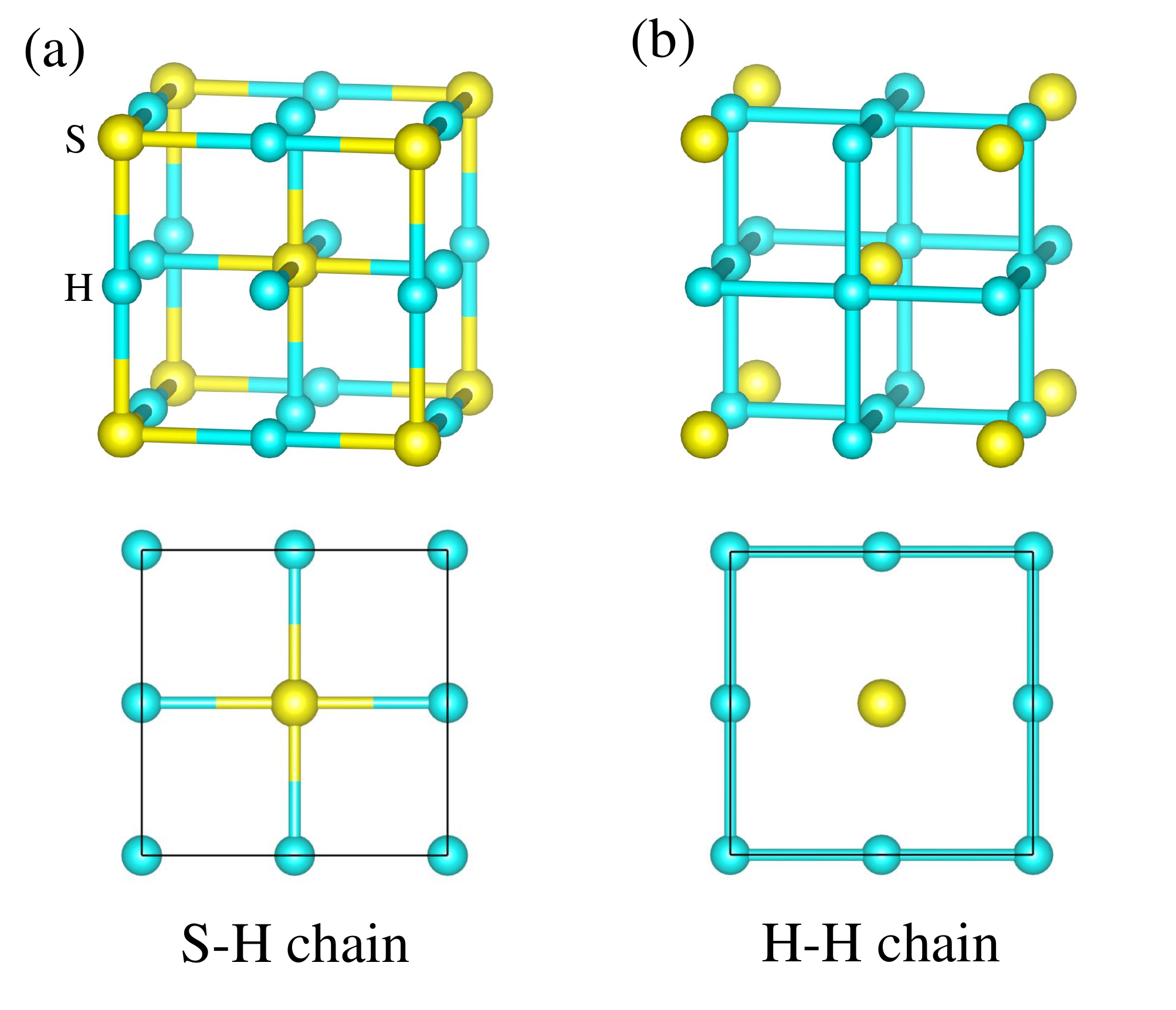}
\caption{ Oblique view (upper panel) and main view (lower panel) of the crystal structures of SH$_3$, illustrating the linkage of (a) S-H bonds and (b) H-H bonds. }
\label{Fig_SH3_structure}
\end{figure}

 The crystal structure of SH$_3$ at high pressure is illustrated in Figs. \ref{Fig_SH3_structure}(a) and \ref{Fig_SH3_structure}(b), where S and H atoms form a cubic lattice. In the conventional cubic cell of SH$_3$, S atoms occupy the vertices and body-centered sites, whereas H atoms occupy edge- and face-centered sites. The electronic states around the Fermi level are well-represented by seven atomic orbital bases in the primitive unit cell of SH$_3$, including one 3$s$ atomic orbital and three 3$p$ atomic orbitals ($p_x$, $p_y$, $p_z$) originating from the S atom, and three 1$s$ atomic orbitals originating from the three H atoms, respectively. In SH$_3$, each S atom is surrounded by six nearest-neighbor H atoms, while each H atom has two nearest-neighbor S atoms and four nearest-neighbor H atoms. This gives rise to two types of $\sigma$ covalent bonds in SH$_3$: one formed between the $s$ and $p$ orbitals of S and the $s$ orbitals of H along the S-H atomic chain, as depicted in Fig. \ref{Fig_SH3_structure}(a), and the other between two nearest H atomic $s$ orbitals along the H-H chain, as illustrated in Fig. \ref{Fig_SH3_structure}(b). Both types of covalent bonds exhibit the same bond lengths of 1.49 \AA\ at 200 GPa.

\begin{figure}[tb]
\includegraphics[angle=0,scale=0.4]{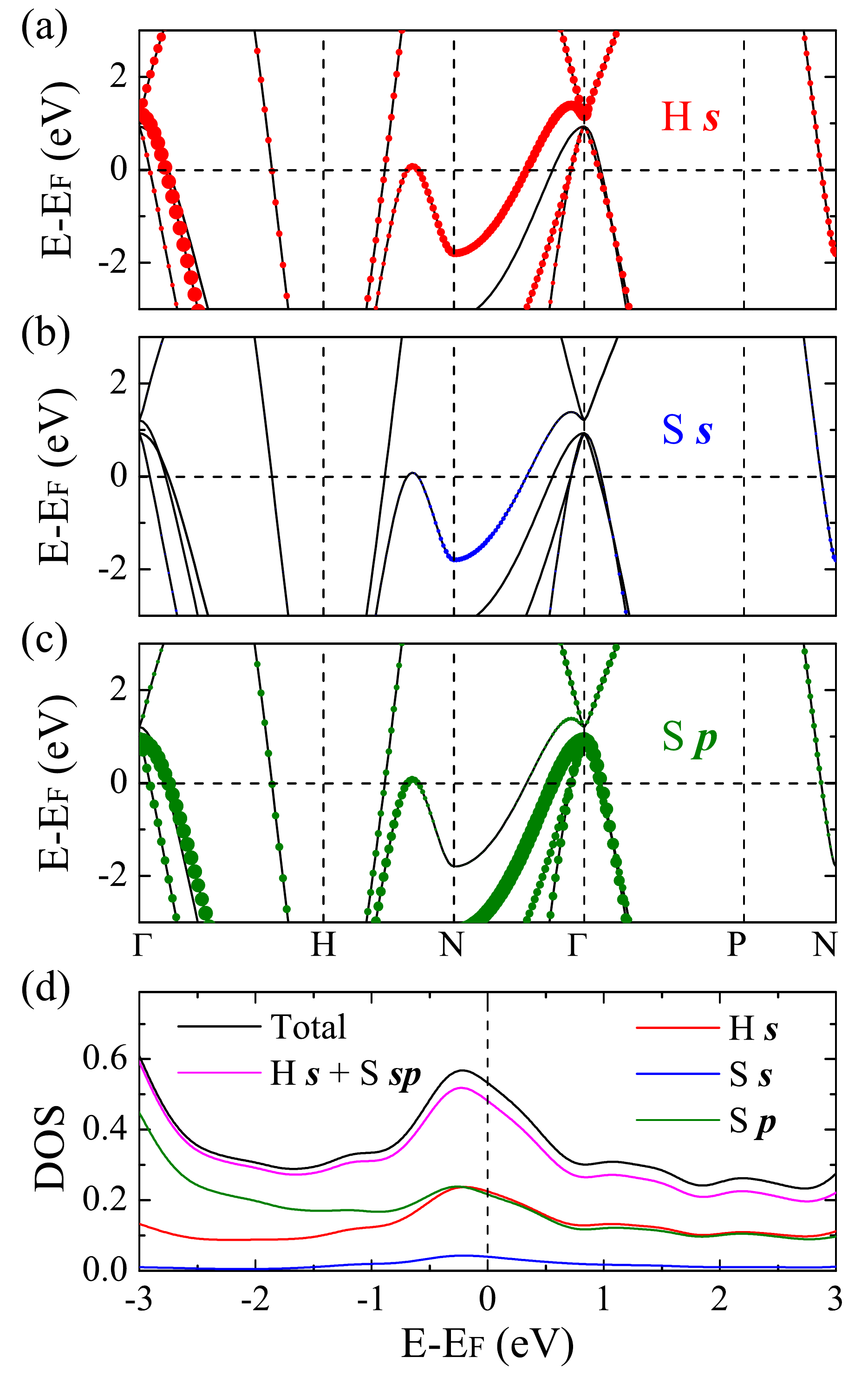}
  \caption{ (a)-(c) Band dispersions of SH$_3$ at 200 GPa, with colored lines indicating the projected weights of (a) H$s$ (red), (b) S$s$ (blue), and (c) S$p$ (green) orbitals, obtained based on the L\"owdin orthogonalization scheme. The widths of the lines are proportional to the weights. (d) Projected partial DOS around the Fermi level is in the unit of state/eV/f.u.. }
\label{Fig_SH3_band1}
\end{figure}

 Figure \ref{Fig_SH3_band1}(a)-(c) shows the spectral weights of three types of orbitals (S$s$, S$p$, and H$s$) around the Fermi level. The corresponding partial DOS is also presented in Fig. \ref{Fig_SH3_band1}(d), revealing the presence of all three types of atomic orbitals around the Fermi level. These three contribute more than 85\% of the electronic DOS around the Fermi level. As shown by the figure, the low-energy bands exhibit significant orbital hybridization. We use COHP to analyze the bonding nature of these hybrid orbitals.

\subsection{Orbital bonding analysis}

\begin{figure}[tb]
\includegraphics[angle=0,scale=0.44]{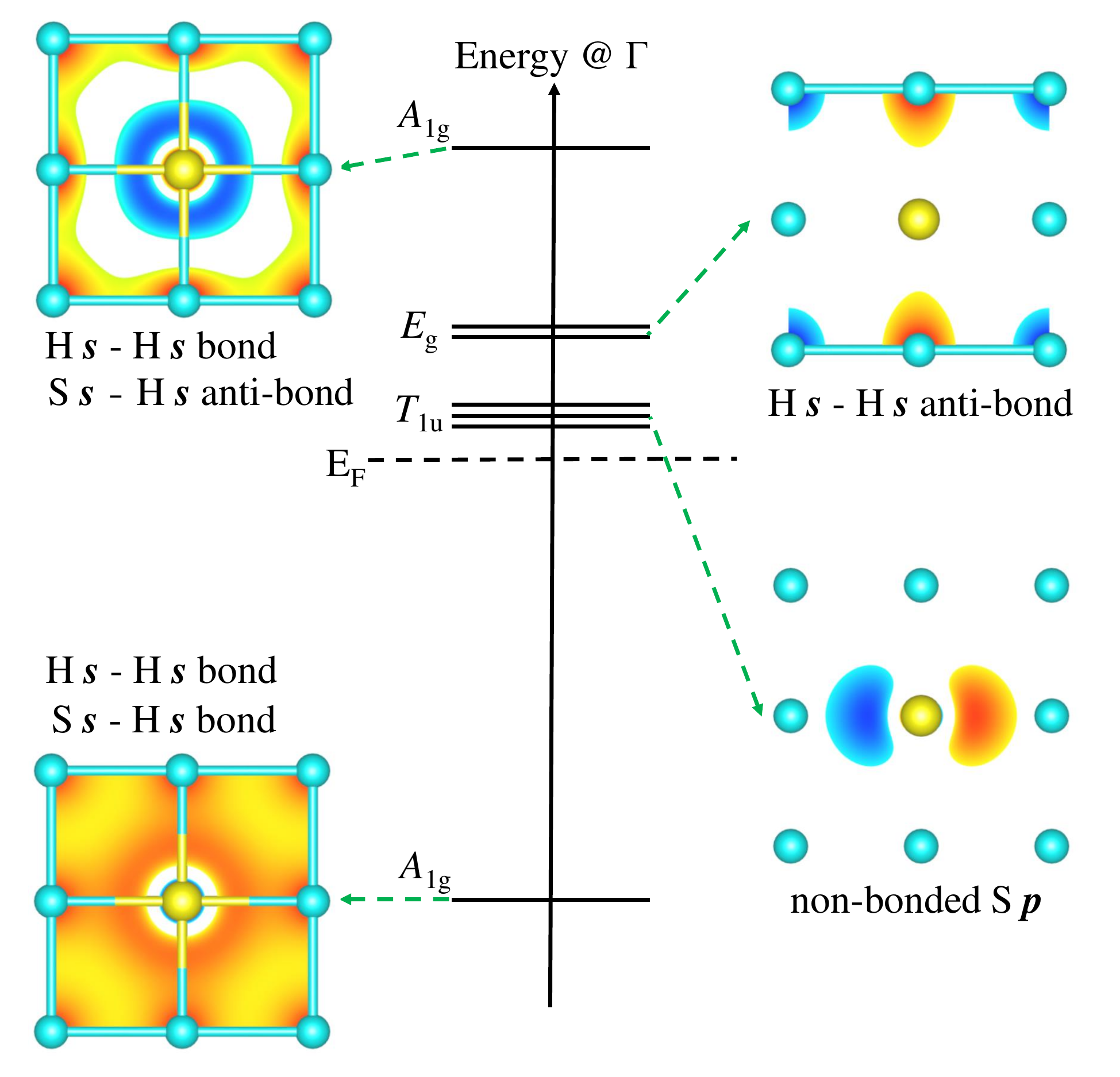}
 \caption{
 Energy levels and their corresponding wave functions of SH$_3$ at the $\Gamma$ point. Red and blue regions represent the orbitals around the atoms for positive and negative phases, respectively. Only the bonding atoms are connected by sticks for each eigenstate.
 }
\label{Fig_SH3_G}
\end{figure}

 In a periodic crystal system, the formation of bonds between two nearest-neighbor atomic orbitals in the Bloch wave function $|\psi_{n{\bf k}}\rangle$ is modulated by the Bloch phase factors. For example, at the $\Gamma$ point, denoted as $|\psi_{n\Gamma}\rangle$, all the translation-equivalent S and H atomic orbitals retain the same phase. Figure \ref{Fig_SH3_G} shows the seven eigenenergies and their corresponding wavefunctions at the $\Gamma$ point. According to the irreducible representation of the $O_\text h$ point group, these seven energy levels contain two distinct $A_\text{1g}$ states, one doubly-degenerate $E_\text g$ state, and one triply-degenerate $T_\text{1u}$ states. Both the $A_\text{1g}$ states correspond to bonding states between neighboring H$s$ orbitals. However, the lower and higher energy $A_{1g}$ correspond to the bonding and antibonding states between the neighboring H$s$ and S$s$ orbitals, respectively. The doubly-degenerate $E_\text g$ states correspond to the antibonding states between two nearest H$s$ orbitals along the H-H chain. The threefold-degenerate $T_\text{1u}$ states originate entirely from the S$p$ orbitals and do not correspond to any bonding or antibonding state between the nearest S or H atoms.

\begin{figure}[t]
\includegraphics[angle=0,scale=0.45]{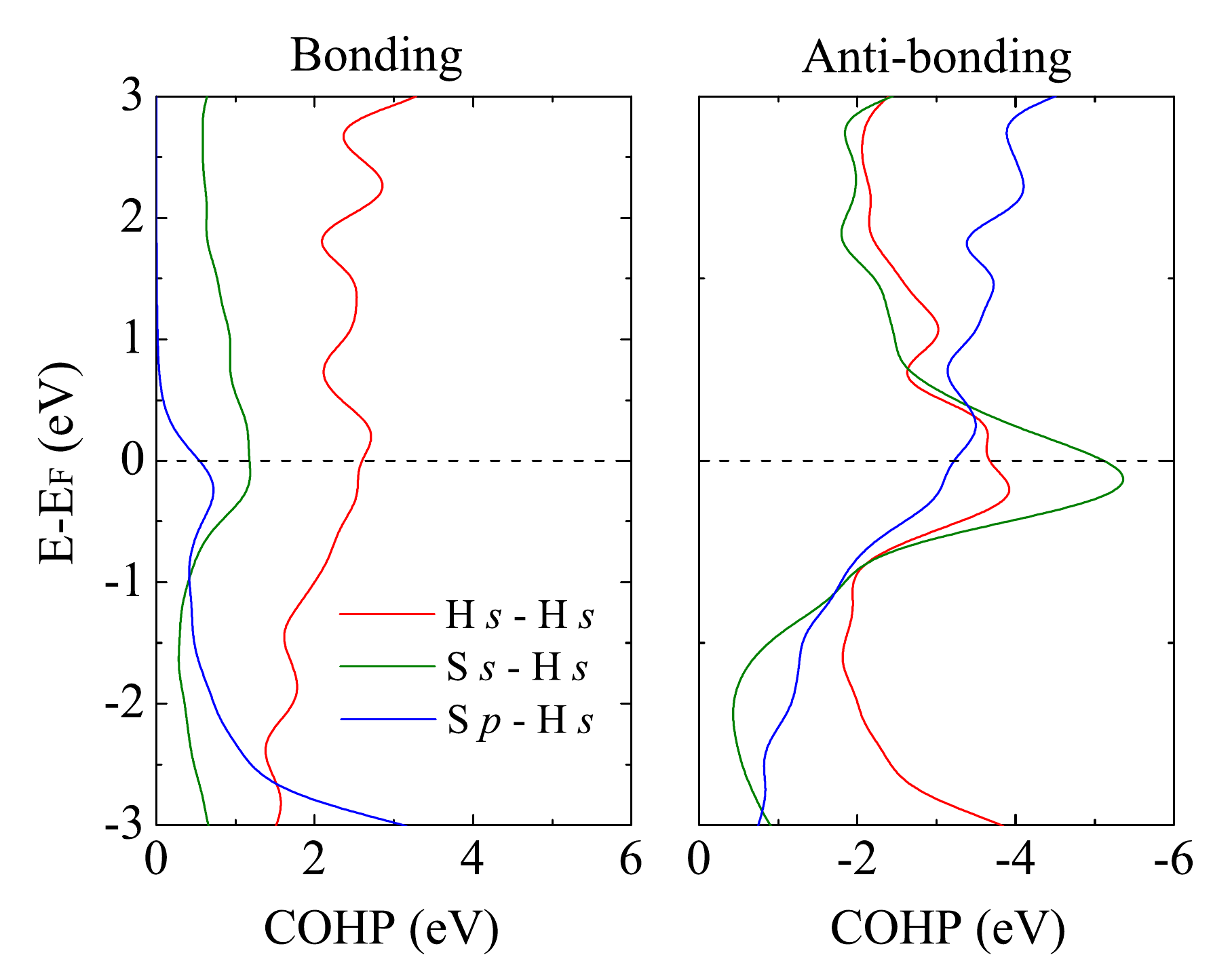}
\caption{
 Distribution of COHP between H$s$-H$s$ (red line), S$p$-H$s$ (blue line), and S$s$-H$s$ (green line) orbitals versus energy. The positive and negative COHP parts correspond to the bonding and antibonding states.}
\label{Fig_SH3_cohp}
\end{figure}

 Now we extend the discussion of the bonding properties of electrons from $\Gamma$ to the whole BZ. Figure \ref{Fig_SH3_band1}(d) shows the orbital-projected DOS obtained via the L\"owdin-orthogonalization scheme. As S$s$, S$p$, and H$s$ orbitals all emerge around the Fermi level, there are six distinct chemical bonds between nearest-neighbor atomic orbitals in SH$_3$. To analyze the relative contributions of these six kinds of chemical bonds to the band structure, we calculated the corresponding COHP in the energy space, as shown in Fig. \ref{Fig_SH3_cohp}. The positive (bonding) and negative (antibonding) parts of COHP were separately integrated to avoid cancellations. Like COHP, the COOP exhibits a similar distribution. They differ from each other only by the ratio between $H_{ij}=\langle i|\hat{H}|j\rangle$ and $S_{ij}=\langle i|j\rangle$.

\begin{table}[b]

\caption{\label{table:hamiltonian_overlap}
 On-site (upper panel) and off-site (middle panel) overlap and energy integrals between the nearest neighbor atomic orbitals. The Fermi level is set as the reference energy $E_0$. Lower panel: the difference between the bonding (antibonding) state eigenenergy, $E_b$ ($E_{ab}$), and the lower (higher) on-site energy $H_{l}$ ($H_{h}$).}
\begin{center}
\begin{tabular*}{8cm}{@{\extracolsep{\fill}} ccc}

\hline \hline
On-site & $S_{ii}$ & $H_{ii}$ (eV) \\
\hline
H $s$ & 1.0 & -18.691 \\
S $p$ & 1.0 & -11.041 \\
S $s$ & 1.0 & -16.467 \\
\hline \hline
Off-site & $S_{ij}$ & $H_{ij}$ (eV) \\
\hline
H $s$ - H $s$ & 0.437 & -11.414 \\
S $p$ - H $s$ & 0.472 &  -9.234 \\
S $s$ - H $s$ & 0.364 & -11.353 \\
\hline \hline
Atomic orbitals & $E_{b} - H_l$ & $E_{ab} - H_{h}$ \\
\hline
H $s$ - H $s$ & -2.259 & 5.766 \\
S $p$ - H $s$ & -0.021 & 2.714 \\
S $s$ - H $s$ & -2.644 & 6.801 \\
\hline\hline
\end{tabular*}
\end{center}
\end{table}

 Table~\ref{table:hamiltonian_overlap} shows the on-site and off-site energies and overlap matrix elements. It should be noted that in a non-orthogonal basis system, the off-diagonal Hamiltonian matrix elements $H_{ij}$ depend on the reference energy $E_0$ chosen, as shown by the equation:
 \begin{equation}
  H_{ij}(E_0') - H_{ij}(E_0) = (E_0 -  E_0') S_{ij}\, .
 \end{equation}
 where $S_{ij}$ is the overlap matrix element between the basis functions $i$ and $j$. Therefore, shifting the reference energy by $(E_0 - E_0')$ results in a change in the Hamiltonian matrix elements by $(E_0 - E_0')S_{ij}$. For the data presented in Table~\ref{table:hamiltonian_overlap}, we set $E_0$ equal to the Fermi energy. The off-site Hamiltonian matrix elements between any two nearest-neighboring atomic orbitals are larger than their on-site energy differences. Specifically, the off-site energies of the H$s$/H$s$ and S$s$/H$s$ orbitals are significantly larger than their on-site energy difference, indicating that both H$s$ - H$s$ and S$s$ - H$s$ $\sigma$ bonds are strong chemical bonds.

 In contrast, the polarized S$p$ - H$s$ $\sigma$ bond is the weakest among them. For two neighbour atomic orbitals, we can characterize their bonding (antibonding) strength by the difference between their hybrid bonding (antibonding) state eigenenergy, $E_b$ ($E_{ab}$), and the lower (higher) atomic orbital on-site energy $H_{l}$ ($H_{h}$). The results are also shown in Table~\ref{table:hamiltonian_overlap}. For the S$p$ - H$s$ bonding state, its energy is very close to that of the nonbonding H$s$ orbital (-0.021 eV), indicating that the bond between S$p$ and H$s$ orbitals is more similar to an ionic bond than a covalent bond.

 \begin{figure}[tb]
\includegraphics[angle=0,scale=0.33]{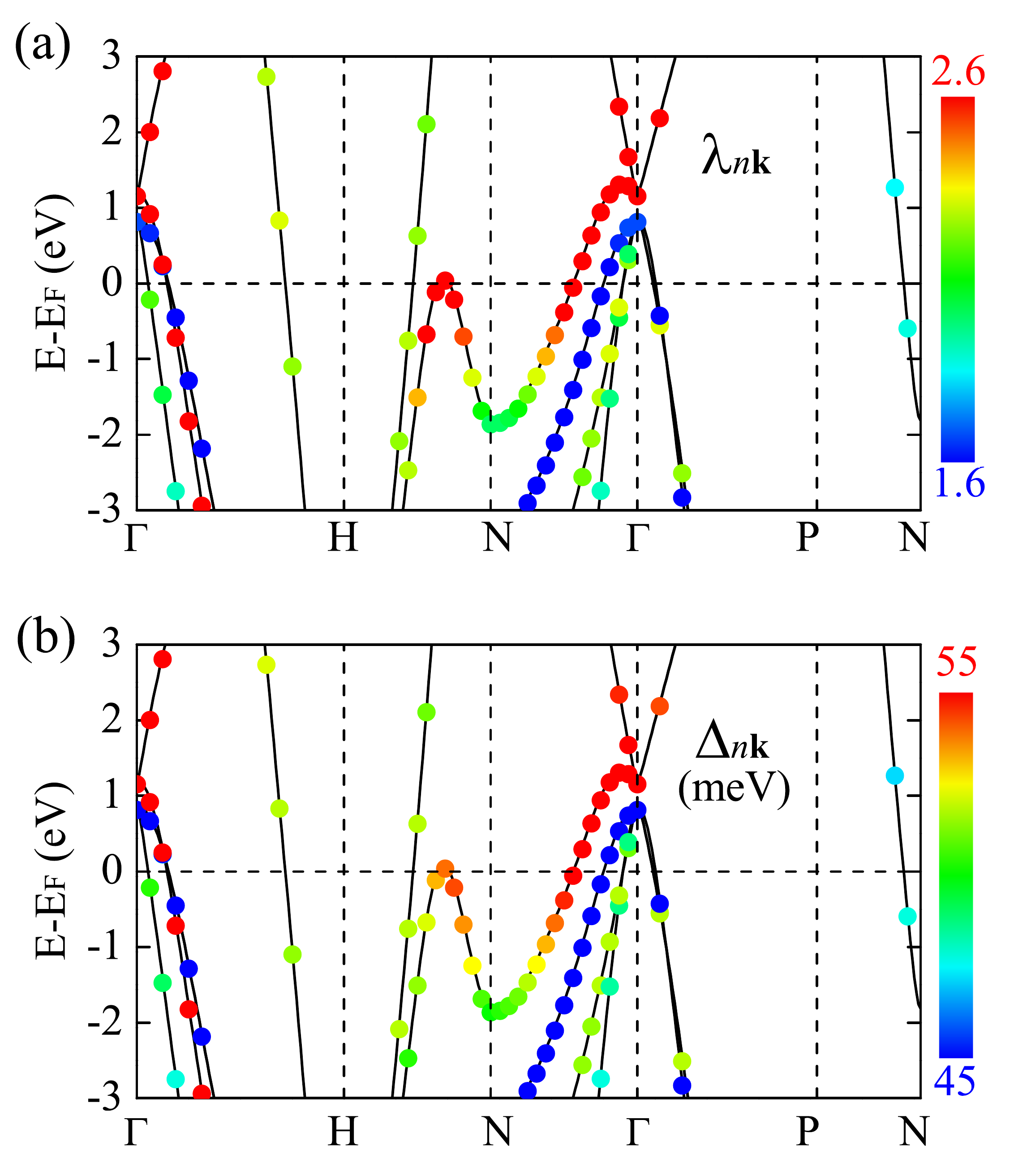}
 \caption{Band structures of SH$_3$ at 200 GPa  near the Fermi level. The Color represents (a) the EPC strength $\lambda_{n{\bf k}}$ and (b) the superconducting gap parameter $\Delta_{n{\bf k}}$ at 20 K. }
\label{Fig_SH3_band2}
\end{figure}

\subsection{Superconducting gap function}

\begin{figure*}[tb]
\includegraphics[angle=0,scale=0.9]{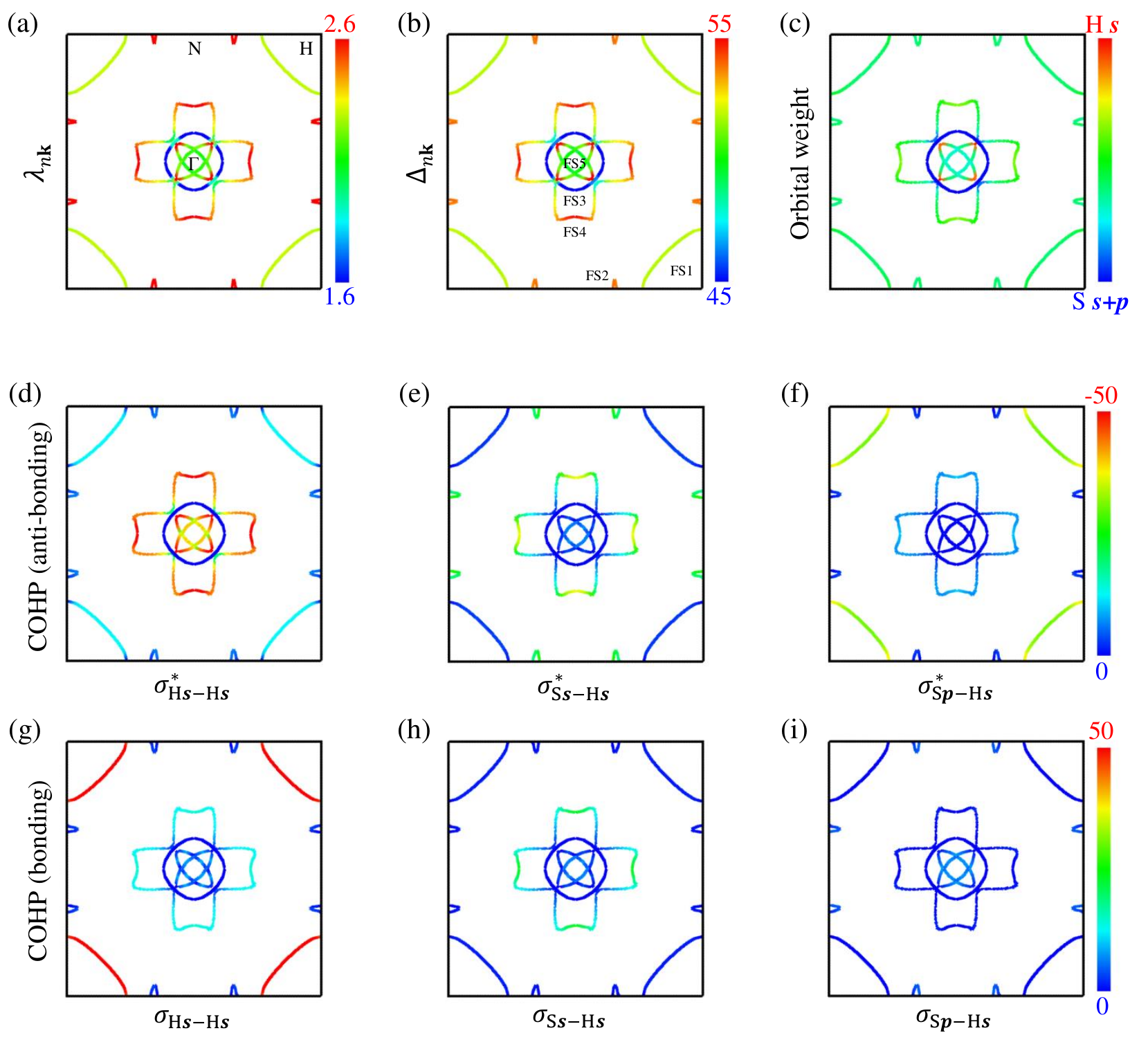}
 \caption{
 Distributions of different physical quantities on the Fermi surfaces: (a) $\lambda_{n{\bf k}}$; (b) $\Delta_{n\mathbf k}$; (c) relative contributions of H$s$ and S $s+p$ orbitals to the electronic states; (d)-(f), the COHP distribution of the antibonding (d) H$s$-H$s$, (e) S$s$-H$s$, and (f) S$p$-H$s$ states; (g)-(i), the COHP distribution of the bonding (g) H$s$-H$s$, (h) S$s$-H$s$, and (i) S$p$-H$s$ states.
 }
\label{Fig_SH3_FS}
\end{figure*}

 It is widely recognized that the superconducting pairing in MgB$_2$ arises from the metalized B-B $\sigma$ bonding states around the Fermi level, as supported by various experimental studies~\cite{mgb2-pcs, mgb2-stm, mgb2-arpes, mgb2-calnature}. Similarly, in SH$_3$, different $\sigma$ and $\sigma^*$ bonds around the Fermi level may also result in superconducting pairing. To investigate this possibility, we calculated the band- and ${\bf k}$-dependent EPC  $\lambda_{n{\bf k}}$ defined by
 \begin{equation}
 \lambda_{n{\bf k}} = \frac{1}{N(E_F)} \sum_{n'{\bf k'}} \lambda({n{\bf k},n'{\bf k'}},0) \delta (E_{n'\bf k'}-E_F),
\label{eq:lambda_nk}
\end{equation}
 and the superconducting gap function $\Delta_{n{\bf k}}$ around the Fermi level by solving the anisotropic Migdal-Eliashberg equation~\cite{ani-ME}. Here, $\lambda_{n{\bf k}}$ measures the strength of the electronic state $|n{\bf k}\rangle$ scattered by phonons.

 Figure~\ref{Fig_SH3_band2} presents the wave vector dependence of $\lambda_{n \mathbf k}$ and $\Delta_{n \mathbf k}$ for SH$_3$. In contrast to MgB$_2$, which is a quasi-two-dimensional superconductor with distinct two-band superconductivity contributed by the $\sigma$- and $\pi$-bonding electrons, SH$_3$ is essentially a three-dimensional material. In SH$_3$, as a single H$s$ orbital can meanwhile bond with all its neighbor atom orbitals, different kinds of bonds strongly couple with each other and result in a smooth evolution of $\lambda_{n{\mathbf k}}$ on the Fermi surface. Consequently, the multiband character of superconductivity is less pronounced in SH$_3$. Moreover, due to the entanglement of different types of chemical bonds in SH$_3$, it is challenging to discern their contributions to EPC.

 In Figure~\ref{Fig_SH3_FS}, we illustrate the Fermi surfaces of SH$_3$ in a representative two-dimensional Brillouin zone (BZ) cut (the $N-\Gamma -H$ plane). Five types of Fermi surfaces are identified: FS1 comprises electron-type Fermi pockets around $H$ and its equivalent points; FS2 consists of multiple hole-type tiny pockets arising from the van Hove singularity along the $N-H$ line; FS3 is a hole-type Fermi surface around $\Gamma$; and FS4 and FS5 are two cross-shaped Fermi surfaces, outside and inside FS3, respectively. As demonstrated in Fig.~\ref{Fig_SH3_FS}(a)-(b), the strongest EPC $\lambda_{n{\bf k}}$, similarly the superconducting energy gap $\Delta_{n\mathbf k}$, are contributed by the electronic states on FS2, FS4, and FS5, while the weakest EPC occurs on FS3.

 By comparing the superconducting gap function $\Delta_{n\mathbf k}$ [Fig. \ref{Fig_SH3_FS}(b)] with the orbital weights [Fig. \ref{Fig_SH3_FS}(c)] and the bonding populations, shown in Fig.~\ref{Fig_SH3_FS}(d)-(i), we find that there is a strong connection between the metalized $\sigma$-antibonding states and the strong pairing function on the Fermi surface. Specifically, the pronounced EPC in the cross-shaped FS4 and FS5 pockets is contributed mainly by the $\sigma^*_{{\text H}s-{\text H}s}$ electrons [Fig.~\ref{Fig_SH3_FS}(d)], whose COHP distribution aligns with the strength variation of $\Delta_{n\bf k}$ on FS4 and FS5. Although other bonding states, including $\sigma_{{\text H}s-{\text H}s}$ [Fig.\ref{Fig_SH3_FS}(g)], $\sigma_{{\text S}s-{\text H}s}$ [Fig.\ref{Fig_SH3_FS}(h)], and $\sigma^*_{{\text S}s-{\text H}s}$ [Fig.~\ref{Fig_SH3_FS}(e)] states, also contribute to EPC on FS4, their contributions are minor compared to that of $\sigma^*_{{\text H}s-{\text H}s}$. Furthermore, the strong EPC observed in the small FS2 pockets is primarily contributed by the $\sigma^*_{{\text S}s-{\text H}s}$ states [Fig.\ref{Fig_SH3_FS}(e)].

 The Fermi surface FS3 displays the weakest EPC and the weakest bonding between nearest-neighboring atoms in SH$_3$. Our investigation of each orbital's contribution to the spectral weight at the $\Gamma$ point reveals that the band associated with FS3 is predominantly formed by non-bonding S$p$ orbitals, specifically the $T_{\text {1u}}$ orbitals depicted in Fig. \ref{Fig_SH3_G}. On the other hand, FS1 primarily arises from the $\sigma_{{\text H}s-{\text H}s}$ [Fig.\ref{Fig_SH3_FS}(g)] and $\sigma^*_{{\text S}p-{\text H}s}$ [Fig.\ref{Fig_SH3_FS}(f)] states, which induce only moderate EPC.

 Among the bonds listed in Table~\ref{table:hamiltonian_overlap}, the bonding strength of $\sigma^*_{{\text S}s-{\text H}s}$ is the strongest. Previous theoretical studies~\cite{SH3vhs1,SH3vhs2} have attributed the van Hove singularities associated with these bonding states as a critical factor for high-$T_c$ superconductivity in SH$_3$. However, our study suggests that the metalized $\sigma^*_{{\text H}s-{\text H}s}$ states, as the second strongest in Table~\ref{table:hamiltonian_overlap}, may play a more significant role in pairing electrons, as demonstrated by comparing their COHP distributions with those of the $\sigma^*_{{\text S}s-{\text H}s}$ states on the three Fermi surfaces with strong EPC: FS2, FS4, and FS5.

\subsection{Electron-phonon coupling}

\begin{figure}[tb]
\includegraphics[angle=0,scale=0.52]{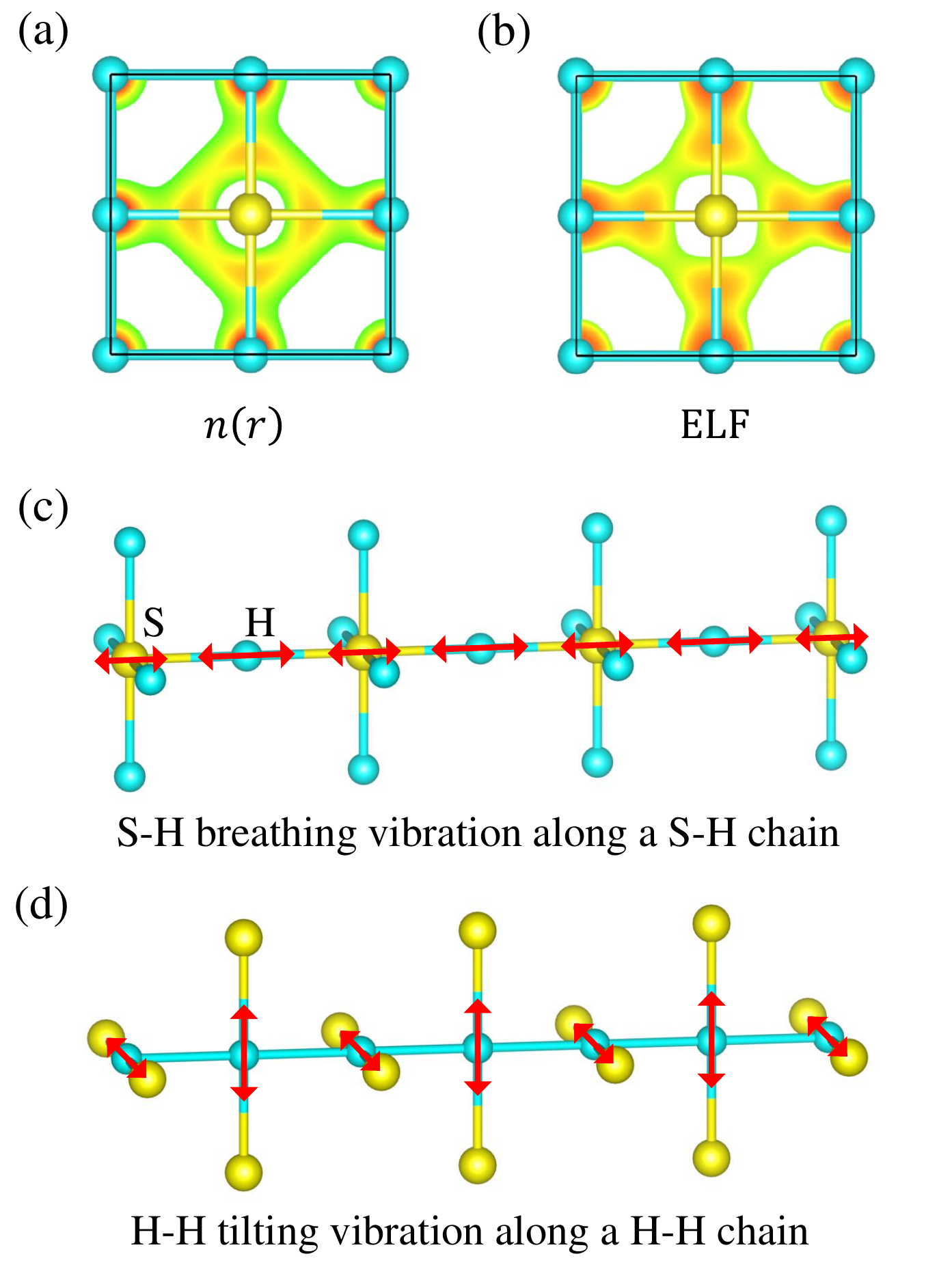}
\caption{
 (a) Charge density distributions $n(\mathbf r)$ and (b) electron localization function (ELF) of the electronic states around the Fermi level between E$_\text F$-0.1 eV and E$_\text F$+0.1 eV of SH$_3$ at 200 GPa.  (c) The S-H breathing vibration along the S-H chain. (d) The H-H tilting vibration in which H vibrates along the direction linking the neighboring H and S atoms on the H-H chain.
 }
\label{Fig_SH3_chgph}
\end{figure}

\begin{figure}[tb]
\includegraphics[angle=0,scale=0.50]{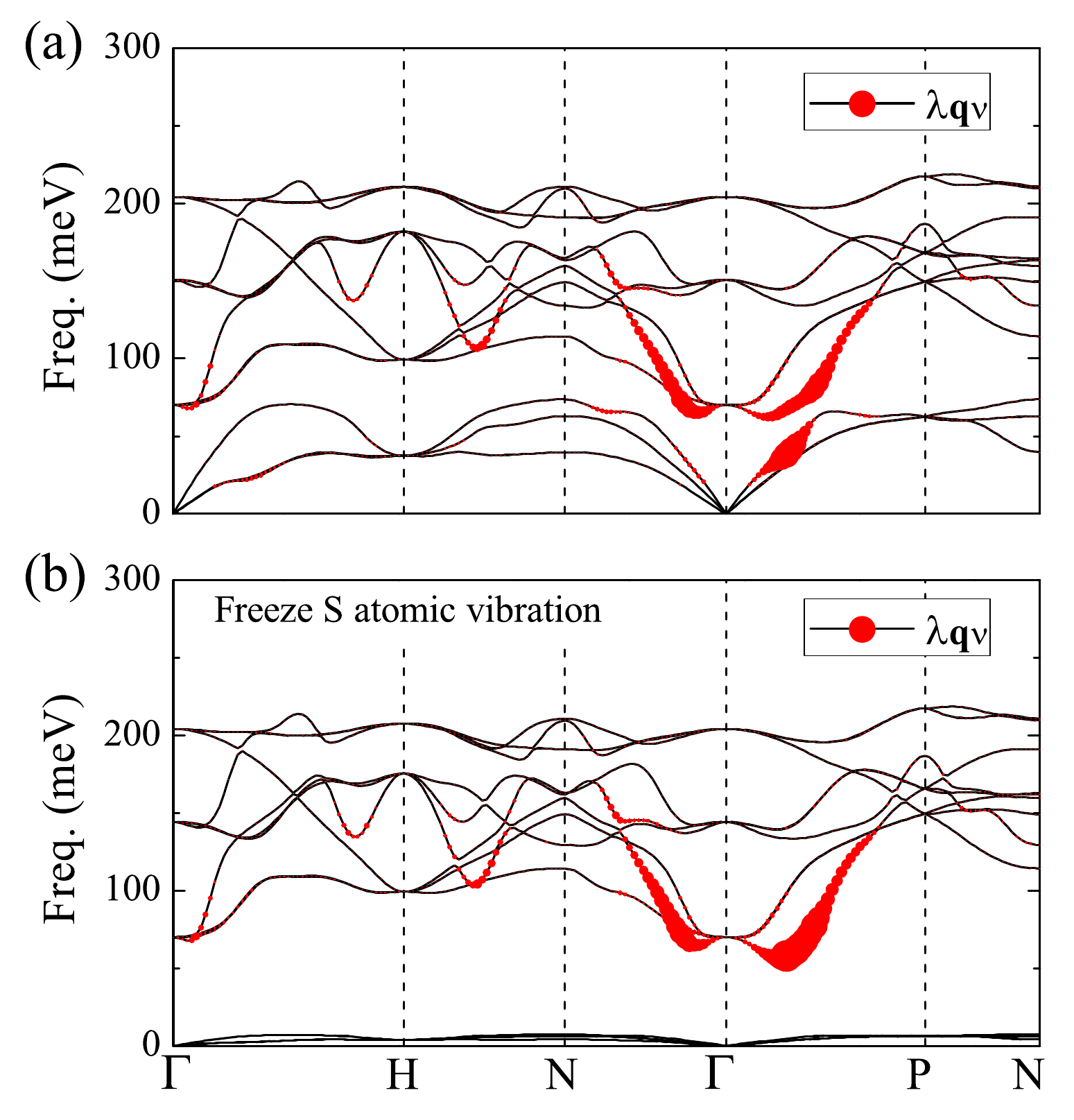}
\caption{
   (a) The phonon spectra of SH$_3$ at 200 GPa. The sizes of the red dots are proportional to $\lambda_{\bf q\nu}$. (b) Same as for panel (a), but the S atomic vibrations are partially frozen by increasing the S mass by 100 times.
 }
\label{Fig_SH3_Sfix}
\end{figure}

 In MgB$_2$, the B-B $\sigma$ bonding orbitals predominantly distribute between the bonding B atoms and are highly susceptible to their relative breathing vibrations. Based on the DFPT~\cite{dfptreview,dfptreview2} calculations, the atomic-vibration induced EPC matrix element, the charge density or self-consistent potential variation, and the wave-function perturbation show a positive correlation with each other. Thus, it is commonly believed that the strong EPC in MgB$_2$ is primarily due to the coupling between its metalized B-B $\sigma$ bond and the B stretching vibration~\cite{MgB2_sigma,mgb2-calnature,mgb2-ph}.

 By examining the charge distribution of SH$_3$ at high pressure, we find that similar properties exist in SH$_3$. Figure~\ref{Fig_SH3_chgph}(a) shows the real-space local density of states $n({\bf r})$ of the electronic states around the Fermi level, whose electrons tend to accumulate along the S-H chain. Similar properties can also be observed in the electron localization function (ELF), shown in panel (b), indicating that there are indeed electrons occupying the S-H $\sigma$ bond.

  Figure \ref{Fig_SH3_Sfix}(a) depicts the phonon spectra of SH$_3$ at 200 GPa. The red dots in the figure represent the EPC strength $\lambda_{\bf q\nu}$, which depends on the phonon mode. At the $\Gamma$ point, the lowest optical modes exhibit weak EPC. In contrast, the phonon modes along the $\Gamma$-N and $\Gamma$-P directions of the Brillouin zone display a significant EPC-induced softening and the strongest EPC. These vibration modes primarily originate from the H atomic vibration along the S-H direction. This characteristic is further emphasized when considering phonon anharmonicity in the calculation~\cite{SH3vhs1,SH3anh1,SH3anh2}. Including phonon anharmonicity leads to a pronounced hardening of the frequency of phonon modes with H atomic vibration along the S-H chain, effectively separating them from the S atomic vibrations. The EPC calculation also reveals that the acoustic modes dominated by the S atom contribute only one-third of the total EPC constant ($\lambda_{\text {tot}}$ = 1.84)~\cite{SH3anh1}.

  These results suggest that the primary contribution to EPC comes from the H atomic vibration along the S-H direction. At first glance, one might assume that this coupling mainly arises from the S-H stretched vibration mode involving the metalized S-H $\sigma$ bond. However, it should be noted that the H atomic vibration along the S-H atomic chain do not only participate in the breathing vibration with S atoms along the S-H chain, as shown in Fig.\ref{Fig_SH3_chgph}(c), but also in the tilting vibration modes involving other H atoms along the H-H chain, as shown in Fig.\ref{Fig_SH3_chgph}(d).

  To distinguish the contribution of the H-S breathing mode to $\lambda_{\bf q\nu}$ from that of the H-H tilting mode, we repeated an EPC calculation where the S atomic vibration was partially frozen by increasing its nucleus mass by 100 times. The resulting EPC spectrum is presented in Fig. ~\ref{Fig_SH3_Sfix}(b). Upon partially freezing the S atomic vibration, we find that the acoustic phonon frequency is dramatically suppressed, indicating that the S atomic vibration predominantly governs the acoustic phonon mode. Conversely, the spectra of the optical phonons remained relatively unchanged when S atoms were partially frozen, and the EPC strength of these optical modes, particularly those along the $\Gamma$-N and $\Gamma$-P directions, exhibits similar behavior as depicted in Fig. ~\ref{Fig_SH3_chgph}(c).

  These observations suggest that the optical phonon modes primarily result from H atomic vibrations. Nevertheless, S atoms still play a vital role in driving SH$_3$ into a high-T$_c$ superconductor. The S atoms provide not only an internal chemical pressure to release the external pressure needed to turn the compound into a superconductor but also an effective charging potential to lift the chemical potential so that the H$s$ $\sigma$-antibonding band can be metalized.

  Based on the above discussion, it is evident that the strong EPC in the optical phonon modes primarily arises from the H-H tilting vibration rather than the S-H breathing vibration. Considering that over 2/3 of the total EPC constant $\lambda_\mathrm{tot}$ originates from the contribution of the optical phonon modes~\cite{SH3anh1}, we  conclude that the strong EPC in SH$_3$ primarily results from the interaction of antibonding H$s$ electrons around the Fermi level with the H-H tilting vibrations.

\subsection{Hydrogen $\sigma$-antibonding states}

\begin{figure}[tb]
\includegraphics[angle=0,scale=0.4]{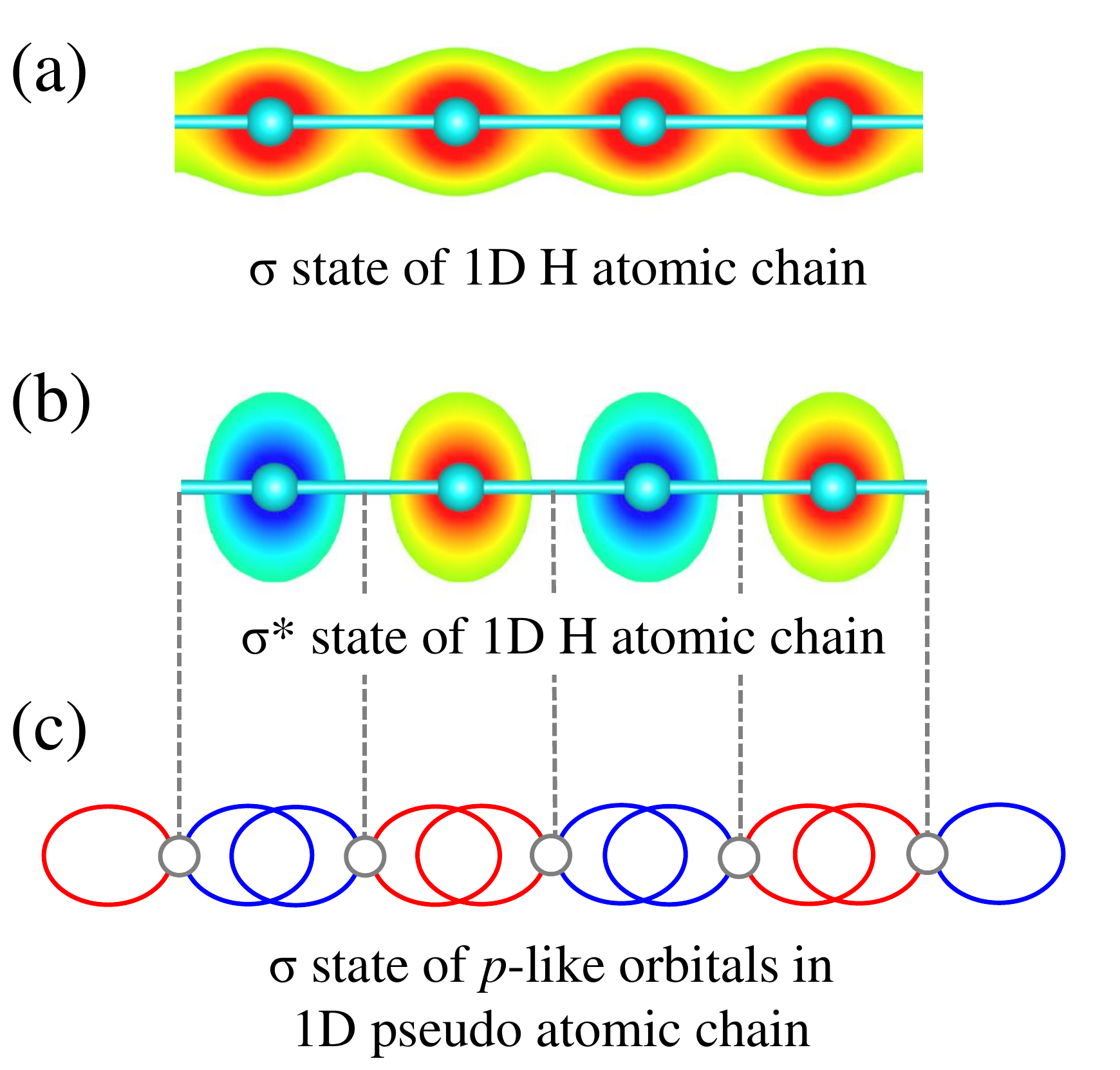}
 \caption{Illustration of the wave functions of (a) H$s$ $\sigma$-bonding and (b) H$s$ $\sigma$-antibonding states along the H-H chain. The red and blue circles represent the positive and negative phases, respectively. The H$s$ chain of the $\sigma$-antibonding orbitals, shown in (b), can also be understood as a chain of pseudo-atoms with ``$p$-like'' bonding orbitals, shown in (c). The pseudo-atom is located at the midpoint of two neighboring H atoms. }
\label{Fig_SH3_dis}
\end{figure}

  In an H $\sigma$-bonding state, electrons accumulate around the midpoint between two H atoms [Fig. \ref{Fig_SH3_dis}(a)]. However, in an H $\sigma$-antibonding state, electrons diffuse away from the midpoint between two H atoms [Fig. \ref{Fig_SH3_dis}(b)]. Hence the low-energy charge distribution would vanish around the midpoint between two neighboring $H$ atoms. This agrees with the numerical result of the low-energy charge distribution shown in Fig.~\ref{Fig_SH3_chgph}(a). This peculiar feature of the $\sigma$-antibonding state is often overlooked in the analysis of low-energy charge distribution, leading to the image that the S-H breathing mode dominates EPC in SH$_3$.

  In a lattice system, electrons in the H$s$ $\sigma$-antibonding states tend to accumulate on the atomic sites and keep away from the midpoints of neighboring H atoms. In the H atomic network of SH$_3$, every H atom flows in two H-H atomic chains with orthometric directions. Therefore, the H$s$ $\sigma$-antibonding states will be pushed away from these two directions and extend to the third direction.    This distribution property makes the tilting vibration of H atoms, namely the vibration along the third direction, can most effectively perturb the $\sigma^*_{{\text H}s-{\text H}s}$ states around the Fermi level. Since this tilting vibration points precisely along the same direction as the S-H breathing vibration, one may inadvertently count its contribution to the pairing interaction on coupling electrons with the S-H breathing modes.

 In a hydrogen molecule, a $\sigma$-antibonding state is a dissolving force to bond a pair of electrons rather than a glue force. However, in a lattice system, an H$s$ $\sigma$-antibonding state can be viewed as an effective $\sigma$-bonding state formed by two $p$-like orbitals whose centers are located at the midpoints of H atoms, as illustrated by Fig.~ \ref{Fig_SH3_dis}(c). This explains why the H$s$ $\sigma$-antibonding electrons can bear strong EPC and contribute to the high-T$_c$ pairing once metalized upon electron doping under high pressure. We believe this picture holds not only for SH$_3$, but also for other hydrogen-rich high-T$_c$ superconductors under ultrahigh pressure.

\section{Summary}

  Through non-orthogonal atomic orbital-based electronic structure calculations, this study investigated the chemical bonding properties between neighboring S/H atoms in SH$_3$ at 200 GPa. Our COOP/COHP analyses revealed that the electronic states around the Fermi level contain both S-H and H-H covalent bonding or antibonding states. Furthermore, by solving the anisotropic Migdal-Eliashberg equation, we found that the EPC strength $\lambda_{n{\bf k}}$ and superconducting gap $\Delta_{n{\bf k}}$ exhibit substantial anisotropy on the Fermi surfaces.

  By comparing the bonding populations and anisotropic superconducting properties, we found that the H-H antibonding states extend widely across the Fermi surfaces and exhibit strong coupling with optical phonon modes generated by the tilting vibrations of hydrogen atoms. This coupling contributes substantially to both $\lambda_{n{\bf k}}$ and $\Delta_{n{\bf k}}$. Conversely, the acoustic phonon modes are predominantly associated with the vibrations of S atoms. Specifically, the coupling between electrons and the strong S-H breathing mode only influences the EPC in the acoustic phonon branches.

  In total, our investigation shows that the metalized H$s$ antibonding electrons contribute more significantly to the total EPC constant through their coupling with the tilting vibration modes of H atoms, compared to the metalized S-H covalent bonding and antibonding electrons in SH$_3$. This discovery highlights an unnoticed yet crucial role played by the metalized H-H antibonding electrons in the high-T$_c$ pairing of SH$_3$. It complements and alters the existing picture that high-T$_c$ pairing primarily arises from metalized S-H covalent bonds in this material. We believe the physical picture revealed in this work holds true more generally and  provides valuable insight into the  understanding of high-T$_c$ superconducting in other hydrogen-rich compounds under ultrahigh pressure.

\begin{acknowledgments}
 This work was supported by the National Natural Science Foundation of China (No.11888101), the National Key Research and Development Program of China (Grant No. 2017YFA0302900), and the Project funded by China Postdoctoral Science Foundation (No. 2022M723355).
\end{acknowledgments}

\bibliography{SH3}

\begin{thebibliography}{59}%
\makeatletter
\providecommand \@ifxundefined [1]{%
 \@ifx{#1\undefined}
}%
\providecommand \@ifnum [1]{%
 \ifnum #1\expandafter \@firstoftwo
 \else \expandafter \@secondoftwo
 \fi
}%
\providecommand \@ifx [1]{%
 \ifx #1\expandafter \@firstoftwo
 \else \expandafter \@secondoftwo
 \fi
}%
\providecommand \natexlab [1]{#1}%
\providecommand \enquote  [1]{``#1''}%
\providecommand \bibnamefont  [1]{#1}%
\providecommand \bibfnamefont [1]{#1}%
\providecommand \citenamefont [1]{#1}%
\providecommand \href@noop [0]{\@secondoftwo}%
\providecommand \href [0]{\begingroup \@sanitize@url \@href}%
\providecommand \@href[1]{\@@startlink{#1}\@@href}%
\providecommand \@@href[1]{\endgroup#1\@@endlink}%
\providecommand \@sanitize@url [0]{\catcode `\\12\catcode `\$12\catcode
  `\&12\catcode `\#12\catcode `\^12\catcode `\_12\catcode `\%12\relax}%
\providecommand \@@startlink[1]{}%
\providecommand \@@endlink[0]{}%
\providecommand \url  [0]{\begingroup\@sanitize@url \@url }%
\providecommand \@url [1]{\endgroup\@href {#1}{\urlprefix }}%
\providecommand \urlprefix  [0]{URL }%
\providecommand \Eprint [0]{\href }%
\providecommand \doibase [0]{https://doi.org/}%
\providecommand \selectlanguage [0]{\@gobble}%
\providecommand \bibinfo  [0]{\@secondoftwo}%
\providecommand \bibfield  [0]{\@secondoftwo}%
\providecommand \translation [1]{[#1]}%
\providecommand \BibitemOpen [0]{}%
\providecommand \bibitemStop [0]{}%
\providecommand \bibitemNoStop [0]{.\EOS\space}%
\providecommand \EOS [0]{\spacefactor3000\relax}%
\providecommand \BibitemShut  [1]{\csname bibitem#1\endcsname}%
\let\auto@bib@innerbib\@empty
\bibitem [{\citenamefont {Dagotto}(1994)}]{cup1}%
  \BibitemOpen
  \bibfield  {author} {\bibinfo {author} {\bibfnamefont {E.}~\bibnamefont
  {Dagotto}},\ }\bibfield  {title} {\bibinfo {title} {Correlated electrons in
  high-temperature superconductors},\ }\href
  {https://doi.org/10.1103/RevModPhys.66.763} {\bibfield  {journal} {\bibinfo
  {journal} {Rev. Mod. Phys.}\ }\textbf {\bibinfo {volume} {66}},\ \bibinfo
  {pages} {763} (\bibinfo {year} {1994})}\BibitemShut {NoStop}%
\bibitem [{\citenamefont {Damascelli}\ \emph {et~al.}(2003)\citenamefont
  {Damascelli}, \citenamefont {Hussain},\ and\ \citenamefont {Shen}}]{cup2}%
  \BibitemOpen
  \bibfield  {author} {\bibinfo {author} {\bibfnamefont {A.}~\bibnamefont
  {Damascelli}}, \bibinfo {author} {\bibfnamefont {Z.}~\bibnamefont
  {Hussain}},\ and\ \bibinfo {author} {\bibfnamefont {Z.-X.}\ \bibnamefont
  {Shen}},\ }\bibfield  {title} {\bibinfo {title} {Angle-resolved photoemission
  studies of the cuprate superconductors},\ }\href
  {https://doi.org/10.1103/RevModPhys.75.473} {\bibfield  {journal} {\bibinfo
  {journal} {Rev. Mod. Phys.}\ }\textbf {\bibinfo {volume} {75}},\ \bibinfo
  {pages} {473} (\bibinfo {year} {2003})}\BibitemShut {NoStop}%
\bibitem [{\citenamefont {Lee}\ \emph {et~al.}(2006)\citenamefont {Lee},
  \citenamefont {Nagaosa},\ and\ \citenamefont {Wen}}]{cup3}%
  \BibitemOpen
  \bibfield  {author} {\bibinfo {author} {\bibfnamefont {P.~A.}\ \bibnamefont
  {Lee}}, \bibinfo {author} {\bibfnamefont {N.}~\bibnamefont {Nagaosa}},\ and\
  \bibinfo {author} {\bibfnamefont {X.-G.}\ \bibnamefont {Wen}},\ }\bibfield
  {title} {\bibinfo {title} {Doping a mott insulator: Physics of
  high-temperature superconductivity},\ }\href
  {https://doi.org/10.1103/RevModPhys.78.17} {\bibfield  {journal} {\bibinfo
  {journal} {Rev. Mod. Phys.}\ }\textbf {\bibinfo {volume} {78}},\ \bibinfo
  {pages} {17} (\bibinfo {year} {2006})}\BibitemShut {NoStop}%
\bibitem [{\citenamefont {Micnas}\ \emph {et~al.}(1990)\citenamefont {Micnas},
  \citenamefont {Ranninger},\ and\ \citenamefont {Robaszkiewicz}}]{cup4}%
  \BibitemOpen
  \bibfield  {author} {\bibinfo {author} {\bibfnamefont {R.}~\bibnamefont
  {Micnas}}, \bibinfo {author} {\bibfnamefont {J.}~\bibnamefont {Ranninger}},\
  and\ \bibinfo {author} {\bibfnamefont {S.}~\bibnamefont {Robaszkiewicz}},\
  }\bibfield  {title} {\bibinfo {title} {Superconductivity in narrow-band
  systems with local nonretarded attractive interactions},\ }\href
  {https://doi.org/10.1103/RevModPhys.62.113} {\bibfield  {journal} {\bibinfo
  {journal} {Rev. Mod. Phys.}\ }\textbf {\bibinfo {volume} {62}},\ \bibinfo
  {pages} {113} (\bibinfo {year} {1990})}\BibitemShut {NoStop}%
\bibitem [{\citenamefont {Kamihara}\ \emph {et~al.}(2008)\citenamefont
  {Kamihara}, \citenamefont {Watanabe}, \citenamefont {Hirano},\ and\
  \citenamefont {Hosono}}]{iro1}%
  \BibitemOpen
  \bibfield  {author} {\bibinfo {author} {\bibfnamefont {Y.}~\bibnamefont
  {Kamihara}}, \bibinfo {author} {\bibfnamefont {T.}~\bibnamefont {Watanabe}},
  \bibinfo {author} {\bibfnamefont {M.}~\bibnamefont {Hirano}},\ and\ \bibinfo
  {author} {\bibfnamefont {H.}~\bibnamefont {Hosono}},\ }\bibfield  {title}
  {\bibinfo {title} {Iron-based layered superconductor $\text
  {LaO}_{1-x}\text{F}_x\text{FeAs}$ ($x$ = 0.05-0.12) with \text{T}$_c$ = 26
  \text{K}},\ }\href {https://doi.org/10.1021/ja800073m} {\bibfield  {journal}
  {\bibinfo  {journal} {J. Am. Chem. Soc.}\ }\textbf {\bibinfo {volume}
  {130}},\ \bibinfo {pages} {3296} (\bibinfo {year} {2008})}\BibitemShut
  {NoStop}%
\bibitem [{\citenamefont {Ma}\ \emph {et~al.}(2008)\citenamefont {Ma},
  \citenamefont {Lu},\ and\ \citenamefont {Xiang}}]{iro2}%
  \BibitemOpen
  \bibfield  {author} {\bibinfo {author} {\bibfnamefont {F.}~\bibnamefont
  {Ma}}, \bibinfo {author} {\bibfnamefont {Z.-Y.}\ \bibnamefont {Lu}},\ and\
  \bibinfo {author} {\bibfnamefont {T.}~\bibnamefont {Xiang}},\ }\bibfield
  {title} {\bibinfo {title} {Arsenic-bridged antiferromagnetic superexchange
  interactions in lafeaso},\ }\href
  {https://doi.org/10.1103/PhysRevB.78.224517} {\bibfield  {journal} {\bibinfo
  {journal} {Phys. Rev. B}\ }\textbf {\bibinfo {volume} {78}},\ \bibinfo
  {pages} {224517} (\bibinfo {year} {2008})}\BibitemShut {NoStop}%
\bibitem [{\citenamefont {Rotter}\ \emph {et~al.}(2008)\citenamefont {Rotter},
  \citenamefont {Tegel},\ and\ \citenamefont {Johrendt}}]{iro3}%
  \BibitemOpen
  \bibfield  {author} {\bibinfo {author} {\bibfnamefont {M.}~\bibnamefont
  {Rotter}}, \bibinfo {author} {\bibfnamefont {M.}~\bibnamefont {Tegel}},\ and\
  \bibinfo {author} {\bibfnamefont {D.}~\bibnamefont {Johrendt}},\ }\bibfield
  {title} {\bibinfo {title} {Superconductivity at 38 \text{K} in the \text{Iron
  Arsenide} $(\text{Ba}_{1-x}\text{K}_{x})\text{Fe}_2\text{As}_2$},\ }\href
  {https://doi.org/10.1103/PhysRevLett.101.107006} {\bibfield  {journal}
  {\bibinfo  {journal} {Phys. Rev. Lett.}\ }\textbf {\bibinfo {volume} {101}},\
  \bibinfo {pages} {107006} (\bibinfo {year} {2008})}\BibitemShut {NoStop}%
\bibitem [{\citenamefont {Hsu}\ \emph {et~al.}(2008)\citenamefont {Hsu},
  \citenamefont {Luo}, \citenamefont {Yeh}, \citenamefont {Chen}, \citenamefont
  {Huang}, \citenamefont {Wu}, \citenamefont {Lee}, \citenamefont {Huang},
  \citenamefont {Chu}, \citenamefont {Yan},\ and\ \citenamefont {Wu}}]{iro4}%
  \BibitemOpen
  \bibfield  {author} {\bibinfo {author} {\bibfnamefont {F.-C.}\ \bibnamefont
  {Hsu}}, \bibinfo {author} {\bibfnamefont {J.-Y.}\ \bibnamefont {Luo}},
  \bibinfo {author} {\bibfnamefont {K.-W.}\ \bibnamefont {Yeh}}, \bibinfo
  {author} {\bibfnamefont {T.-K.}\ \bibnamefont {Chen}}, \bibinfo {author}
  {\bibfnamefont {T.-W.}\ \bibnamefont {Huang}}, \bibinfo {author}
  {\bibfnamefont {P.~M.}\ \bibnamefont {Wu}}, \bibinfo {author} {\bibfnamefont
  {Y.-C.}\ \bibnamefont {Lee}}, \bibinfo {author} {\bibfnamefont {Y.-L.}\
  \bibnamefont {Huang}}, \bibinfo {author} {\bibfnamefont {Y.-Y.}\ \bibnamefont
  {Chu}}, \bibinfo {author} {\bibfnamefont {D.-C.}\ \bibnamefont {Yan}},\ and\
  \bibinfo {author} {\bibfnamefont {M.-K.}\ \bibnamefont {Wu}},\ }\bibfield
  {title} {\bibinfo {title} {Superconductivity in the \text{PbO}-type structure
  $\alpha$-\text{FeSe}},\ }\href {https://doi.org/10.1073/pnas.0807325105}
  {\bibfield  {journal} {\bibinfo  {journal} {Proc. Natl. Acad. Sci.}\ }\textbf
  {\bibinfo {volume} {105}},\ \bibinfo {pages} {14262} (\bibinfo {year}
  {2008})}\BibitemShut {NoStop}%
\bibitem [{\citenamefont {Alireza}\ \emph {et~al.}(2008)\citenamefont
  {Alireza}, \citenamefont {Ko}, \citenamefont {Gillett}, \citenamefont
  {Petrone}, \citenamefont {Cole}, \citenamefont {Lonzarich},\ and\
  \citenamefont {Sebastian}}]{iro5}%
  \BibitemOpen
  \bibfield  {author} {\bibinfo {author} {\bibfnamefont {P.~L.}\ \bibnamefont
  {Alireza}}, \bibinfo {author} {\bibfnamefont {Y.~T.~C.}\ \bibnamefont {Ko}},
  \bibinfo {author} {\bibfnamefont {J.}~\bibnamefont {Gillett}}, \bibinfo
  {author} {\bibfnamefont {C.~M.}\ \bibnamefont {Petrone}}, \bibinfo {author}
  {\bibfnamefont {J.~M.}\ \bibnamefont {Cole}}, \bibinfo {author}
  {\bibfnamefont {G.~G.}\ \bibnamefont {Lonzarich}},\ and\ \bibinfo {author}
  {\bibfnamefont {S.~E.}\ \bibnamefont {Sebastian}},\ }\bibfield  {title}
  {\bibinfo {title} {Superconductivity up to 29 \text{K} in
  $\text{SrFe}_2\text{As}_2$ and $\text{BaFe}_2\text{As}_2$ at high
  pressures},\ }\href {https://doi.org/10.1088/0953-8984/21/1/012208}
  {\bibfield  {journal} {\bibinfo  {journal} {J. Phys.: Condens. Matter}\
  }\textbf {\bibinfo {volume} {21}},\ \bibinfo {pages} {012208} (\bibinfo
  {year} {2008})}\BibitemShut {NoStop}%
\bibitem [{\citenamefont {Kimber}\ \emph {et~al.}(2009)\citenamefont {Kimber},
  \citenamefont {Kreyssig}, \citenamefont {Zhang}, \citenamefont {Jeschke},
  \citenamefont {Valent{\'i}}, \citenamefont {Yokaichiya}, \citenamefont
  {Colombier}, \citenamefont {Yan}, \citenamefont {Hansen}, \citenamefont
  {Chatterji}, \citenamefont {McQueeney}, \citenamefont {Canfield},
  \citenamefont {Goldman},\ and\ \citenamefont {Argyriou}}]{iro6}%
  \BibitemOpen
  \bibfield  {author} {\bibinfo {author} {\bibfnamefont {S.~A.~J.}\
  \bibnamefont {Kimber}}, \bibinfo {author} {\bibfnamefont {A.}~\bibnamefont
  {Kreyssig}}, \bibinfo {author} {\bibfnamefont {Y.-Z.}\ \bibnamefont {Zhang}},
  \bibinfo {author} {\bibfnamefont {H.~O.}\ \bibnamefont {Jeschke}}, \bibinfo
  {author} {\bibfnamefont {R.}~\bibnamefont {Valent{\'i}}}, \bibinfo {author}
  {\bibfnamefont {F.}~\bibnamefont {Yokaichiya}}, \bibinfo {author}
  {\bibfnamefont {E.}~\bibnamefont {Colombier}}, \bibinfo {author}
  {\bibfnamefont {J.}~\bibnamefont {Yan}}, \bibinfo {author} {\bibfnamefont
  {T.~C.}\ \bibnamefont {Hansen}}, \bibinfo {author} {\bibfnamefont
  {T.}~\bibnamefont {Chatterji}}, \bibinfo {author} {\bibfnamefont {R.~J.}\
  \bibnamefont {McQueeney}}, \bibinfo {author} {\bibfnamefont {P.~C.}\
  \bibnamefont {Canfield}}, \bibinfo {author} {\bibfnamefont {A.~I.}\
  \bibnamefont {Goldman}},\ and\ \bibinfo {author} {\bibfnamefont {D.~N.}\
  \bibnamefont {Argyriou}},\ }\bibfield  {title} {\bibinfo {title}
  {Similarities between structural distortions under pressure and chemical
  doping in superconducting $\text{BaFe}_2\text{As}_2$},\ }\href
  {https://doi.org/10.1038/nmat2443} {\bibfield  {journal} {\bibinfo  {journal}
  {Nat. Mater.}\ }\textbf {\bibinfo {volume} {8}},\ \bibinfo {pages} {471}
  (\bibinfo {year} {2009})}\BibitemShut {NoStop}%
\bibitem [{\citenamefont {Stewart}(1984)}]{hea1}%
  \BibitemOpen
  \bibfield  {author} {\bibinfo {author} {\bibfnamefont {G.~R.}\ \bibnamefont
  {Stewart}},\ }\bibfield  {title} {\bibinfo {title} {Heavy-fermion systems},\
  }\href {https://doi.org/10.1103/RevModPhys.56.755} {\bibfield  {journal}
  {\bibinfo  {journal} {Rev. Mod. Phys.}\ }\textbf {\bibinfo {volume} {56}},\
  \bibinfo {pages} {755} (\bibinfo {year} {1984})}\BibitemShut {NoStop}%
\bibitem [{\citenamefont {Emery}\ and\ \citenamefont {Kivelson}(1995)}]{hea2}%
  \BibitemOpen
  \bibfield  {author} {\bibinfo {author} {\bibfnamefont {V.~J.}\ \bibnamefont
  {Emery}}\ and\ \bibinfo {author} {\bibfnamefont {S.~A.}\ \bibnamefont
  {Kivelson}},\ }\bibfield  {title} {\bibinfo {title} {Importance of phase
  fluctuations in superconductors with small superfluid density},\ }\href
  {https://doi.org/10.1038/374434a0} {\bibfield  {journal} {\bibinfo  {journal}
  {Nature}\ }\textbf {\bibinfo {volume} {374}},\ \bibinfo {pages} {434}
  (\bibinfo {year} {1995})}\BibitemShut {NoStop}%
\bibitem [{\citenamefont {Mathur}\ \emph {et~al.}(1998)\citenamefont {Mathur},
  \citenamefont {Grosche}, \citenamefont {Julian}, \citenamefont {Walker},
  \citenamefont {Freye}, \citenamefont {Haselwimmer},\ and\ \citenamefont
  {Lonzarich}}]{hea3}%
  \BibitemOpen
  \bibfield  {author} {\bibinfo {author} {\bibfnamefont {N.~D.}\ \bibnamefont
  {Mathur}}, \bibinfo {author} {\bibfnamefont {F.~M.}\ \bibnamefont {Grosche}},
  \bibinfo {author} {\bibfnamefont {S.~R.}\ \bibnamefont {Julian}}, \bibinfo
  {author} {\bibfnamefont {I.~R.}\ \bibnamefont {Walker}}, \bibinfo {author}
  {\bibfnamefont {D.~M.}\ \bibnamefont {Freye}}, \bibinfo {author}
  {\bibfnamefont {R.~K.~W.}\ \bibnamefont {Haselwimmer}},\ and\ \bibinfo
  {author} {\bibfnamefont {G.~G.}\ \bibnamefont {Lonzarich}},\ }\bibfield
  {title} {\bibinfo {title} {Magnetically mediated superconductivity in heavy
  fermion compounds},\ }\href {https://doi.org/10.1038/27838} {\bibfield
  {journal} {\bibinfo  {journal} {Nature}\ }\textbf {\bibinfo {volume} {394}},\
  \bibinfo {pages} {39} (\bibinfo {year} {1998})}\BibitemShut {NoStop}%
\bibitem [{\citenamefont {Sigrist}\ and\ \citenamefont {Ueda}(1991)}]{hea4}%
  \BibitemOpen
  \bibfield  {author} {\bibinfo {author} {\bibfnamefont {M.}~\bibnamefont
  {Sigrist}}\ and\ \bibinfo {author} {\bibfnamefont {K.}~\bibnamefont {Ueda}},\
  }\bibfield  {title} {\bibinfo {title} {Phenomenological theory of
  unconventional superconductivity},\ }\href
  {https://doi.org/10.1103/RevModPhys.63.239} {\bibfield  {journal} {\bibinfo
  {journal} {Rev. Mod. Phys.}\ }\textbf {\bibinfo {volume} {63}},\ \bibinfo
  {pages} {239} (\bibinfo {year} {1991})}\BibitemShut {NoStop}%
\bibitem [{\citenamefont {Blatter}\ \emph {et~al.}(1994)\citenamefont
  {Blatter}, \citenamefont {Feigel'man}, \citenamefont {Geshkenbein},
  \citenamefont {Larkin},\ and\ \citenamefont {Vinokur}}]{spinflu}%
  \BibitemOpen
  \bibfield  {author} {\bibinfo {author} {\bibfnamefont {G.}~\bibnamefont
  {Blatter}}, \bibinfo {author} {\bibfnamefont {M.~V.}\ \bibnamefont
  {Feigel'man}}, \bibinfo {author} {\bibfnamefont {V.~B.}\ \bibnamefont
  {Geshkenbein}}, \bibinfo {author} {\bibfnamefont {A.~I.}\ \bibnamefont
  {Larkin}},\ and\ \bibinfo {author} {\bibfnamefont {V.~M.}\ \bibnamefont
  {Vinokur}},\ }\bibfield  {title} {\bibinfo {title} {Vortices in
  high-temperature superconductors},\ }\href
  {https://doi.org/10.1103/RevModPhys.66.1125} {\bibfield  {journal} {\bibinfo
  {journal} {Rev. Mod. Phys.}\ }\textbf {\bibinfo {volume} {66}},\ \bibinfo
  {pages} {1125} (\bibinfo {year} {1994})}\BibitemShut {NoStop}%
\bibitem [{\citenamefont {Bardeen}\ \emph {et~al.}(1957)\citenamefont
  {Bardeen}, \citenamefont {Cooper},\ and\ \citenamefont {Schrieffer}}]{bcs}%
  \BibitemOpen
  \bibfield  {author} {\bibinfo {author} {\bibfnamefont {J.}~\bibnamefont
  {Bardeen}}, \bibinfo {author} {\bibfnamefont {L.~N.}\ \bibnamefont
  {Cooper}},\ and\ \bibinfo {author} {\bibfnamefont {J.~R.}\ \bibnamefont
  {Schrieffer}},\ }\bibfield  {title} {\bibinfo {title} {Microscopic theory of
  superconductivity},\ }\href {https://doi.org/10.1103/PhysRev.106.162}
  {\bibfield  {journal} {\bibinfo  {journal} {Phys. Rev.}\ }\textbf {\bibinfo
  {volume} {106}},\ \bibinfo {pages} {162} (\bibinfo {year}
  {1957})}\BibitemShut {NoStop}%
\bibitem [{\citenamefont {Ashcroft}(1968)}]{metalHpre}%
  \BibitemOpen
  \bibfield  {author} {\bibinfo {author} {\bibfnamefont {N.~W.}\ \bibnamefont
  {Ashcroft}},\ }\bibfield  {title} {\bibinfo {title} {Metallic hydrogen: A
  high-temperature superconductor?},\ }\href
  {https://doi.org/10.1103/PhysRevLett.21.1748} {\bibfield  {journal} {\bibinfo
   {journal} {Phys. Rev. Lett.}\ }\textbf {\bibinfo {volume} {21}},\ \bibinfo
  {pages} {1748} (\bibinfo {year} {1968})}\BibitemShut {NoStop}%
\bibitem [{\citenamefont {Zhang}\ \emph {et~al.}(2007)\citenamefont {Zhang},
  \citenamefont {Niu}, \citenamefont {Li}, \citenamefont {Cui}, \citenamefont
  {Wang}, \citenamefont {Ma}, \citenamefont {He},\ and\ \citenamefont
  {Zou}}]{metalHcal}%
  \BibitemOpen
  \bibfield  {author} {\bibinfo {author} {\bibfnamefont {L.}~\bibnamefont
  {Zhang}}, \bibinfo {author} {\bibfnamefont {Y.}~\bibnamefont {Niu}}, \bibinfo
  {author} {\bibfnamefont {Q.}~\bibnamefont {Li}}, \bibinfo {author}
  {\bibfnamefont {T.}~\bibnamefont {Cui}}, \bibinfo {author} {\bibfnamefont
  {Y.}~\bibnamefont {Wang}}, \bibinfo {author} {\bibfnamefont {Y.}~\bibnamefont
  {Ma}}, \bibinfo {author} {\bibfnamefont {Z.}~\bibnamefont {He}},\ and\
  \bibinfo {author} {\bibfnamefont {G.}~\bibnamefont {Zou}},\ }\bibfield
  {title} {\bibinfo {title} {Ab initio prediction of superconductivity in
  molecular metallic hydrogen under high pressure},\ }\href
  {https://doi.org/https://doi.org/10.1016/j.ssc.2006.12.029} {\bibfield
  {journal} {\bibinfo  {journal} {Solid State Commun.}\ }\textbf {\bibinfo
  {volume} {141}},\ \bibinfo {pages} {610} (\bibinfo {year}
  {2007})}\BibitemShut {NoStop}%
\bibitem [{\citenamefont {Dubrovinsky}\ \emph {et~al.}(2012)\citenamefont
  {Dubrovinsky}, \citenamefont {Dubrovinskaia}, \citenamefont {Prakapenka},\
  and\ \citenamefont {Abakumov}}]{presstec}%
  \BibitemOpen
  \bibfield  {author} {\bibinfo {author} {\bibfnamefont {L.}~\bibnamefont
  {Dubrovinsky}}, \bibinfo {author} {\bibfnamefont {N.}~\bibnamefont
  {Dubrovinskaia}}, \bibinfo {author} {\bibfnamefont {V.~B.}\ \bibnamefont
  {Prakapenka}},\ and\ \bibinfo {author} {\bibfnamefont {A.~M.}\ \bibnamefont
  {Abakumov}},\ }\bibfield  {title} {\bibinfo {title} {Implementation of
  micro-ball nanodiamond anvils for high-pressure studies above 6
  \text{Mbar}},\ }\href {https://doi.org/10.1038/ncomms2160} {\bibfield
  {journal} {\bibinfo  {journal} {Nat. Commun.}\ }\textbf {\bibinfo {volume}
  {3}},\ \bibinfo {pages} {1163} (\bibinfo {year} {2012})}\BibitemShut
  {NoStop}%
\bibitem [{\citenamefont {Gao}\ \emph {et~al.}(2015{\natexlab{a}})\citenamefont
  {Gao}, \citenamefont {Lu},\ and\ \citenamefont {Xiang}}]{gaoprb2015}%
  \BibitemOpen
  \bibfield  {author} {\bibinfo {author} {\bibfnamefont {M.}~\bibnamefont
  {Gao}}, \bibinfo {author} {\bibfnamefont {Z.-Y.}\ \bibnamefont {Lu}},\ and\
  \bibinfo {author} {\bibfnamefont {T.}~\bibnamefont {Xiang}},\ }\bibfield
  {title} {\bibinfo {title} {Prediction of phonon-mediated high-temperature
  superconductivity in \text{Li}$_3$\text{B}$_4$\text{C}$_2$},\ }\href
  {https://doi.org/10.1103/PhysRevB.91.045132} {\bibfield  {journal} {\bibinfo
  {journal} {Phys. Rev. B}\ }\textbf {\bibinfo {volume} {91}},\ \bibinfo
  {pages} {045132} (\bibinfo {year} {2015}{\natexlab{a}})}\BibitemShut
  {NoStop}%
\bibitem [{\citenamefont {Gao}\ \emph {et~al.}(2015{\natexlab{b}})\citenamefont
  {Gao}, \citenamefont {Lu},\ and\ \citenamefont {Xiang}}]{gao2015}%
  \BibitemOpen
  \bibfield  {author} {\bibinfo {author} {\bibfnamefont {M.}~\bibnamefont
  {Gao}}, \bibinfo {author} {\bibfnamefont {Z.-Y.}\ \bibnamefont {Lu}},\ and\
  \bibinfo {author} {\bibfnamefont {T.}~\bibnamefont {Xiang}},\ }\bibfield
  {title} {\bibinfo {title} {Finding high-temperature superconductors by
  metallizing the $\sigma$-bonding electrons},\ }\href
  {https://doi.org/10.7693/wl20150701} {\bibfield  {journal} {\bibinfo
  {journal} {Physics}\ }\textbf {\bibinfo {volume} {44}},\ \bibinfo {pages}
  {421} (\bibinfo {year} {2015}{\natexlab{b}})},\ \bibinfo {note} {(in
  Chinese)}\BibitemShut {NoStop}%
\bibitem [{\citenamefont {Nagamatsu}\ \emph {et~al.}(2001)\citenamefont
  {Nagamatsu}, \citenamefont {Nakagawa}, \citenamefont {Muranaka},
  \citenamefont {Zenitani},\ and\ \citenamefont {Akimitsu}}]{MgB2_1st}%
  \BibitemOpen
  \bibfield  {author} {\bibinfo {author} {\bibfnamefont {J.}~\bibnamefont
  {Nagamatsu}}, \bibinfo {author} {\bibfnamefont {N.}~\bibnamefont {Nakagawa}},
  \bibinfo {author} {\bibfnamefont {T.}~\bibnamefont {Muranaka}}, \bibinfo
  {author} {\bibfnamefont {Y.}~\bibnamefont {Zenitani}},\ and\ \bibinfo
  {author} {\bibfnamefont {J.}~\bibnamefont {Akimitsu}},\ }\bibfield  {title}
  {\bibinfo {title} {Superconductivity at 39 \text{K} in magnesium diboride},\
  }\href {https://doi.org/10.1038/35065039} {\bibfield  {journal} {\bibinfo
  {journal} {Nature}\ }\textbf {\bibinfo {volume} {410}},\ \bibinfo {pages}
  {63} (\bibinfo {year} {2001})}\BibitemShut {NoStop}%
\bibitem [{\citenamefont {Drozdov}\ \emph {et~al.}(2015)\citenamefont
  {Drozdov}, \citenamefont {Eremets}, \citenamefont {Troyan}, \citenamefont
  {Ksenofontov},\ and\ \citenamefont {Shylin}}]{SH3_1st}%
  \BibitemOpen
  \bibfield  {author} {\bibinfo {author} {\bibfnamefont {A.~P.}\ \bibnamefont
  {Drozdov}}, \bibinfo {author} {\bibfnamefont {M.~I.}\ \bibnamefont
  {Eremets}}, \bibinfo {author} {\bibfnamefont {I.~A.}\ \bibnamefont {Troyan}},
  \bibinfo {author} {\bibfnamefont {V.}~\bibnamefont {Ksenofontov}},\ and\
  \bibinfo {author} {\bibfnamefont {S.~I.}\ \bibnamefont {Shylin}},\ }\bibfield
   {title} {\bibinfo {title} {Conventional superconductivity at 203 kelvin at
  high pressures in the sulfur hydride system},\ }\href
  {https://doi.org/10.1038/nature14964} {\bibfield  {journal} {\bibinfo
  {journal} {Nature}\ }\textbf {\bibinfo {volume} {525}},\ \bibinfo {pages}
  {73} (\bibinfo {year} {2015})}\BibitemShut {NoStop}%
\bibitem [{\citenamefont {Somayazulu}\ \emph {et~al.}(2019)\citenamefont
  {Somayazulu}, \citenamefont {Ahart}, \citenamefont {Mishra}, \citenamefont
  {Geballe}, \citenamefont {Baldini}, \citenamefont {Meng}, \citenamefont
  {Struzhkin},\ and\ \citenamefont {Hemley}}]{LaH10_1st}%
  \BibitemOpen
  \bibfield  {author} {\bibinfo {author} {\bibfnamefont {M.}~\bibnamefont
  {Somayazulu}}, \bibinfo {author} {\bibfnamefont {M.}~\bibnamefont {Ahart}},
  \bibinfo {author} {\bibfnamefont {A.~K.}\ \bibnamefont {Mishra}}, \bibinfo
  {author} {\bibfnamefont {Z.~M.}\ \bibnamefont {Geballe}}, \bibinfo {author}
  {\bibfnamefont {M.}~\bibnamefont {Baldini}}, \bibinfo {author} {\bibfnamefont
  {Y.}~\bibnamefont {Meng}}, \bibinfo {author} {\bibfnamefont {V.~V.}\
  \bibnamefont {Struzhkin}},\ and\ \bibinfo {author} {\bibfnamefont {R.~J.}\
  \bibnamefont {Hemley}},\ }\bibfield  {title} {\bibinfo {title} {Evidence for
  superconductivity above 260 \text{K} in \text{Lanthanum Superhydride at
  Megabar Pressures}},\ }\href {https://doi.org/10.1103/PhysRevLett.122.027001}
  {\bibfield  {journal} {\bibinfo  {journal} {Phys. Rev. Lett.}\ }\textbf
  {\bibinfo {volume} {122}},\ \bibinfo {pages} {027001} (\bibinfo {year}
  {2019})}\BibitemShut {NoStop}%
\bibitem [{\citenamefont {Drozdov}\ \emph {et~al.}(2019)\citenamefont
  {Drozdov}, \citenamefont {Kong}, \citenamefont {Minkov}, \citenamefont
  {Besedin}, \citenamefont {Kuzovnikov}, \citenamefont {Mozaffari},
  \citenamefont {Balicas}, \citenamefont {Balakirev}, \citenamefont {Graf},
  \citenamefont {Prakapenka}, \citenamefont {Greenberg}, \citenamefont
  {Knyazev}, \citenamefont {Tkacz},\ and\ \citenamefont {Eremets}}]{LaH10_2nd}%
  \BibitemOpen
  \bibfield  {author} {\bibinfo {author} {\bibfnamefont {A.~P.}\ \bibnamefont
  {Drozdov}}, \bibinfo {author} {\bibfnamefont {P.~P.}\ \bibnamefont {Kong}},
  \bibinfo {author} {\bibfnamefont {V.~S.}\ \bibnamefont {Minkov}}, \bibinfo
  {author} {\bibfnamefont {S.~P.}\ \bibnamefont {Besedin}}, \bibinfo {author}
  {\bibfnamefont {M.~A.}\ \bibnamefont {Kuzovnikov}}, \bibinfo {author}
  {\bibfnamefont {S.}~\bibnamefont {Mozaffari}}, \bibinfo {author}
  {\bibfnamefont {L.}~\bibnamefont {Balicas}}, \bibinfo {author} {\bibfnamefont
  {F.~F.}\ \bibnamefont {Balakirev}}, \bibinfo {author} {\bibfnamefont {D.~E.}\
  \bibnamefont {Graf}}, \bibinfo {author} {\bibfnamefont {V.~B.}\ \bibnamefont
  {Prakapenka}}, \bibinfo {author} {\bibfnamefont {E.}~\bibnamefont
  {Greenberg}}, \bibinfo {author} {\bibfnamefont {D.~A.}\ \bibnamefont
  {Knyazev}}, \bibinfo {author} {\bibfnamefont {M.}~\bibnamefont {Tkacz}},\
  and\ \bibinfo {author} {\bibfnamefont {M.~I.}\ \bibnamefont {Eremets}},\
  }\bibfield  {title} {\bibinfo {title} {Superconductivity at 250 \text{K} in
  lanthanum hydride under high pressures},\ }\href
  {https://doi.org/10.1038/s41586-019-1201-8} {\bibfield  {journal} {\bibinfo
  {journal} {Nature}\ }\textbf {\bibinfo {volume} {569}},\ \bibinfo {pages}
  {528} (\bibinfo {year} {2019})}\BibitemShut {NoStop}%
\bibitem [{\citenamefont {Kong}\ \emph {et~al.}(2021)\citenamefont {Kong},
  \citenamefont {Minkov}, \citenamefont {Kuzovnikov}, \citenamefont {Drozdov},
  \citenamefont {Besedin}, \citenamefont {Mozaffari}, \citenamefont {Balicas},
  \citenamefont {Balakirev}, \citenamefont {Prakapenka}, \citenamefont
  {Chariton}, \citenamefont {Knyazev}, \citenamefont {Greenberg},\ and\
  \citenamefont {Eremets}}]{YH9_1st}%
  \BibitemOpen
  \bibfield  {author} {\bibinfo {author} {\bibfnamefont {P.}~\bibnamefont
  {Kong}}, \bibinfo {author} {\bibfnamefont {V.~S.}\ \bibnamefont {Minkov}},
  \bibinfo {author} {\bibfnamefont {M.~A.}\ \bibnamefont {Kuzovnikov}},
  \bibinfo {author} {\bibfnamefont {A.~P.}\ \bibnamefont {Drozdov}}, \bibinfo
  {author} {\bibfnamefont {S.~P.}\ \bibnamefont {Besedin}}, \bibinfo {author}
  {\bibfnamefont {S.}~\bibnamefont {Mozaffari}}, \bibinfo {author}
  {\bibfnamefont {L.}~\bibnamefont {Balicas}}, \bibinfo {author} {\bibfnamefont
  {F.~F.}\ \bibnamefont {Balakirev}}, \bibinfo {author} {\bibfnamefont {V.~B.}\
  \bibnamefont {Prakapenka}}, \bibinfo {author} {\bibfnamefont
  {S.}~\bibnamefont {Chariton}}, \bibinfo {author} {\bibfnamefont {D.~A.}\
  \bibnamefont {Knyazev}}, \bibinfo {author} {\bibfnamefont {E.}~\bibnamefont
  {Greenberg}},\ and\ \bibinfo {author} {\bibfnamefont {M.~I.}\ \bibnamefont
  {Eremets}},\ }\bibfield  {title} {\bibinfo {title} {Superconductivity up to
  243 \text{K} in the yttrium-hydrogen system under high pressure},\ }\href
  {https://doi.org/10.1038/s41467-021-25372-2} {\bibfield  {journal} {\bibinfo
  {journal} {Nat. Commun.}\ }\textbf {\bibinfo {volume} {12}},\ \bibinfo
  {pages} {5075} (\bibinfo {year} {2021})}\BibitemShut {NoStop}%
\bibitem [{\citenamefont {Ma}\ \emph {et~al.}(2022)\citenamefont {Ma},
  \citenamefont {Wang}, \citenamefont {Xie}, \citenamefont {Yang},
  \citenamefont {Wang}, \citenamefont {Zhou}, \citenamefont {Liu},
  \citenamefont {Yu}, \citenamefont {Zhao}, \citenamefont {Wang}, \citenamefont
  {Liu},\ and\ \citenamefont {Ma}}]{CaH6_1st}%
  \BibitemOpen
  \bibfield  {author} {\bibinfo {author} {\bibfnamefont {L.}~\bibnamefont
  {Ma}}, \bibinfo {author} {\bibfnamefont {K.}~\bibnamefont {Wang}}, \bibinfo
  {author} {\bibfnamefont {Y.}~\bibnamefont {Xie}}, \bibinfo {author}
  {\bibfnamefont {X.}~\bibnamefont {Yang}}, \bibinfo {author} {\bibfnamefont
  {Y.}~\bibnamefont {Wang}}, \bibinfo {author} {\bibfnamefont {M.}~\bibnamefont
  {Zhou}}, \bibinfo {author} {\bibfnamefont {H.}~\bibnamefont {Liu}}, \bibinfo
  {author} {\bibfnamefont {X.}~\bibnamefont {Yu}}, \bibinfo {author}
  {\bibfnamefont {Y.}~\bibnamefont {Zhao}}, \bibinfo {author} {\bibfnamefont
  {H.}~\bibnamefont {Wang}}, \bibinfo {author} {\bibfnamefont {G.}~\bibnamefont
  {Liu}},\ and\ \bibinfo {author} {\bibfnamefont {Y.}~\bibnamefont {Ma}},\
  }\bibfield  {title} {\bibinfo {title} {High-temperature superconducting phase
  in clathrate calcium hydride $\text{CaH}_6$ up to 215 \text{K} at a pressure
  of 172 \text{GPa}},\ }\href {https://doi.org/10.1103/PhysRevLett.128.167001}
  {\bibfield  {journal} {\bibinfo  {journal} {Phys. Rev. Lett.}\ }\textbf
  {\bibinfo {volume} {128}},\ \bibinfo {pages} {167001} (\bibinfo {year}
  {2022})}\BibitemShut {NoStop}%
\bibitem [{\citenamefont {Li}\ \emph {et~al.}(2022)\citenamefont {Li},
  \citenamefont {He}, \citenamefont {Zhang}, \citenamefont {Wang},
  \citenamefont {Zhang}, \citenamefont {Jia}, \citenamefont {Feng},
  \citenamefont {Lu}, \citenamefont {Zhao}, \citenamefont {Zhang},
  \citenamefont {Min}, \citenamefont {Long}, \citenamefont {Yu}, \citenamefont
  {Wang}, \citenamefont {Ye}, \citenamefont {Zhang}, \citenamefont
  {Prakapenka}, \citenamefont {Chariton}, \citenamefont {Ginsberg},
  \citenamefont {Bass}, \citenamefont {Yuan}, \citenamefont {Liu},\ and\
  \citenamefont {Jin}}]{CaH6_2nd}%
  \BibitemOpen
  \bibfield  {author} {\bibinfo {author} {\bibfnamefont {Z.}~\bibnamefont
  {Li}}, \bibinfo {author} {\bibfnamefont {X.}~\bibnamefont {He}}, \bibinfo
  {author} {\bibfnamefont {C.}~\bibnamefont {Zhang}}, \bibinfo {author}
  {\bibfnamefont {X.}~\bibnamefont {Wang}}, \bibinfo {author} {\bibfnamefont
  {S.}~\bibnamefont {Zhang}}, \bibinfo {author} {\bibfnamefont
  {Y.}~\bibnamefont {Jia}}, \bibinfo {author} {\bibfnamefont {S.}~\bibnamefont
  {Feng}}, \bibinfo {author} {\bibfnamefont {K.}~\bibnamefont {Lu}}, \bibinfo
  {author} {\bibfnamefont {J.}~\bibnamefont {Zhao}}, \bibinfo {author}
  {\bibfnamefont {J.}~\bibnamefont {Zhang}}, \bibinfo {author} {\bibfnamefont
  {B.}~\bibnamefont {Min}}, \bibinfo {author} {\bibfnamefont {Y.}~\bibnamefont
  {Long}}, \bibinfo {author} {\bibfnamefont {R.}~\bibnamefont {Yu}}, \bibinfo
  {author} {\bibfnamefont {L.}~\bibnamefont {Wang}}, \bibinfo {author}
  {\bibfnamefont {M.}~\bibnamefont {Ye}}, \bibinfo {author} {\bibfnamefont
  {Z.}~\bibnamefont {Zhang}}, \bibinfo {author} {\bibfnamefont
  {V.}~\bibnamefont {Prakapenka}}, \bibinfo {author} {\bibfnamefont
  {S.}~\bibnamefont {Chariton}}, \bibinfo {author} {\bibfnamefont {P.~A.}\
  \bibnamefont {Ginsberg}}, \bibinfo {author} {\bibfnamefont {J.}~\bibnamefont
  {Bass}}, \bibinfo {author} {\bibfnamefont {S.}~\bibnamefont {Yuan}}, \bibinfo
  {author} {\bibfnamefont {H.}~\bibnamefont {Liu}},\ and\ \bibinfo {author}
  {\bibfnamefont {C.}~\bibnamefont {Jin}},\ }\bibfield  {title} {\bibinfo
  {title} {Superconductivity above 200 \text{K} discovered in superhydrides of
  calcium},\ }\href {https://doi.org/10.1038/s41467-022-30454-w} {\bibfield
  {journal} {\bibinfo  {journal} {Nat. Commun.}\ }\textbf {\bibinfo {volume}
  {13}},\ \bibinfo {pages} {2863} (\bibinfo {year} {2022})}\BibitemShut
  {NoStop}%
\bibitem [{\citenamefont {Peng}\ \emph {et~al.}(2017)\citenamefont {Peng},
  \citenamefont {Sun}, \citenamefont {Pickard}, \citenamefont {Needs},
  \citenamefont {Wu},\ and\ \citenamefont {Ma}}]{maprl}%
  \BibitemOpen
  \bibfield  {author} {\bibinfo {author} {\bibfnamefont {F.}~\bibnamefont
  {Peng}}, \bibinfo {author} {\bibfnamefont {Y.}~\bibnamefont {Sun}}, \bibinfo
  {author} {\bibfnamefont {C.~J.}\ \bibnamefont {Pickard}}, \bibinfo {author}
  {\bibfnamefont {R.~J.}\ \bibnamefont {Needs}}, \bibinfo {author}
  {\bibfnamefont {Q.}~\bibnamefont {Wu}},\ and\ \bibinfo {author}
  {\bibfnamefont {Y.}~\bibnamefont {Ma}},\ }\bibfield  {title} {\bibinfo
  {title} {Hydrogen clathrate structures in rare earth hydrides at high
  pressures: Possible route to room-temperature superconductivity},\ }\href
  {https://doi.org/10.1103/PhysRevLett.119.107001} {\bibfield  {journal}
  {\bibinfo  {journal} {Phys. Rev. Lett.}\ }\textbf {\bibinfo {volume} {119}},\
  \bibinfo {pages} {107001} (\bibinfo {year} {2017})}\BibitemShut {NoStop}%
\bibitem [{\citenamefont {Sun}\ \emph {et~al.}(2019)\citenamefont {Sun},
  \citenamefont {Lv}, \citenamefont {Xie}, \citenamefont {Liu},\ and\
  \citenamefont {Ma}}]{maprl2}%
  \BibitemOpen
  \bibfield  {author} {\bibinfo {author} {\bibfnamefont {Y.}~\bibnamefont
  {Sun}}, \bibinfo {author} {\bibfnamefont {J.}~\bibnamefont {Lv}}, \bibinfo
  {author} {\bibfnamefont {Y.}~\bibnamefont {Xie}}, \bibinfo {author}
  {\bibfnamefont {H.}~\bibnamefont {Liu}},\ and\ \bibinfo {author}
  {\bibfnamefont {Y.}~\bibnamefont {Ma}},\ }\bibfield  {title} {\bibinfo
  {title} {Route to a superconducting phase above room temperature in
  electron-doped hydride compounds under high pressure},\ }\href
  {https://doi.org/10.1103/PhysRevLett.123.097001} {\bibfield  {journal}
  {\bibinfo  {journal} {Phys. Rev. Lett.}\ }\textbf {\bibinfo {volume} {123}},\
  \bibinfo {pages} {097001} (\bibinfo {year} {2019})}\BibitemShut {NoStop}%
\bibitem [{\citenamefont {Flores-Livas}\ \emph {et~al.}(2020)\citenamefont
  {Flores-Livas}, \citenamefont {Boeri}, \citenamefont {Sanna}, \citenamefont
  {Profeta}, \citenamefont {Arita},\ and\ \citenamefont
  {Eremets}}]{MH_review2}%
  \BibitemOpen
  \bibfield  {author} {\bibinfo {author} {\bibfnamefont {J.~A.}\ \bibnamefont
  {Flores-Livas}}, \bibinfo {author} {\bibfnamefont {L.}~\bibnamefont {Boeri}},
  \bibinfo {author} {\bibfnamefont {A.}~\bibnamefont {Sanna}}, \bibinfo
  {author} {\bibfnamefont {G.}~\bibnamefont {Profeta}}, \bibinfo {author}
  {\bibfnamefont {R.}~\bibnamefont {Arita}},\ and\ \bibinfo {author}
  {\bibfnamefont {M.}~\bibnamefont {Eremets}},\ }\bibfield  {title} {\bibinfo
  {title} {A perspective on conventional high-temperature superconductors at
  high pressure: Methods and materials},\ }\href
  {https://doi.org/https://doi.org/10.1016/j.physrep.2020.02.003} {\bibfield
  {journal} {\bibinfo  {journal} {Phys. Rep.}\ }\textbf {\bibinfo {volume}
  {856}},\ \bibinfo {pages} {1} (\bibinfo {year} {2020})}\BibitemShut {NoStop}%
\bibitem [{\citenamefont {Duan}\ \emph {et~al.}(2014)\citenamefont {Duan},
  \citenamefont {Liu}, \citenamefont {Tian}, \citenamefont {Li}, \citenamefont
  {Huang}, \citenamefont {Zhao}, \citenamefont {Yu}, \citenamefont {Liu},
  \citenamefont {Tian},\ and\ \citenamefont {Cui}}]{SH3_pre}%
  \BibitemOpen
  \bibfield  {author} {\bibinfo {author} {\bibfnamefont {D.}~\bibnamefont
  {Duan}}, \bibinfo {author} {\bibfnamefont {Y.}~\bibnamefont {Liu}}, \bibinfo
  {author} {\bibfnamefont {F.}~\bibnamefont {Tian}}, \bibinfo {author}
  {\bibfnamefont {D.}~\bibnamefont {Li}}, \bibinfo {author} {\bibfnamefont
  {X.}~\bibnamefont {Huang}}, \bibinfo {author} {\bibfnamefont
  {Z.}~\bibnamefont {Zhao}}, \bibinfo {author} {\bibfnamefont {H.}~\bibnamefont
  {Yu}}, \bibinfo {author} {\bibfnamefont {B.}~\bibnamefont {Liu}}, \bibinfo
  {author} {\bibfnamefont {W.}~\bibnamefont {Tian}},\ and\ \bibinfo {author}
  {\bibfnamefont {T.}~\bibnamefont {Cui}},\ }\bibfield  {title} {\bibinfo
  {title} {Pressure-induced metallization of dense \text{(H2S)2H2} with high-tc
  superconductivity},\ }\href {https://doi.org/10.1038/srep06968} {\bibfield
  {journal} {\bibinfo  {journal} {Sci. Rep.}\ }\textbf {\bibinfo {volume}
  {4}},\ \bibinfo {pages} {6968} (\bibinfo {year} {2014})}\BibitemShut
  {NoStop}%
\bibitem [{\citenamefont {Wang}\ \emph {et~al.}(2012)\citenamefont {Wang},
  \citenamefont {Tse}, \citenamefont {Tanaka}, \citenamefont {Iitaka},\ and\
  \citenamefont {Ma}}]{CaH6_pre}%
  \BibitemOpen
  \bibfield  {author} {\bibinfo {author} {\bibfnamefont {H.}~\bibnamefont
  {Wang}}, \bibinfo {author} {\bibfnamefont {J.~S.}\ \bibnamefont {Tse}},
  \bibinfo {author} {\bibfnamefont {K.}~\bibnamefont {Tanaka}}, \bibinfo
  {author} {\bibfnamefont {T.}~\bibnamefont {Iitaka}},\ and\ \bibinfo {author}
  {\bibfnamefont {Y.}~\bibnamefont {Ma}},\ }\bibfield  {title} {\bibinfo
  {title} {Superconductive sodalite-like clathrate calcium hydride at high
  pressures},\ }\href {https://doi.org/10.1073/pnas.1118168109} {\bibfield
  {journal} {\bibinfo  {journal} {Proc. Natl. Acad. Sci.}\ }\textbf {\bibinfo
  {volume} {109}},\ \bibinfo {pages} {6463} (\bibinfo {year}
  {2012})}\BibitemShut {NoStop}%
\bibitem [{\citenamefont {Bernstein}\ \emph {et~al.}(2015)\citenamefont
  {Bernstein}, \citenamefont {Hellberg}, \citenamefont {Johannes},
  \citenamefont {Mazin},\ and\ \citenamefont {Mehl}}]{SH3str1}%
  \BibitemOpen
  \bibfield  {author} {\bibinfo {author} {\bibfnamefont {N.}~\bibnamefont
  {Bernstein}}, \bibinfo {author} {\bibfnamefont {C.~S.}\ \bibnamefont
  {Hellberg}}, \bibinfo {author} {\bibfnamefont {M.~D.}\ \bibnamefont
  {Johannes}}, \bibinfo {author} {\bibfnamefont {I.~I.}\ \bibnamefont
  {Mazin}},\ and\ \bibinfo {author} {\bibfnamefont {M.~J.}\ \bibnamefont
  {Mehl}},\ }\bibfield  {title} {\bibinfo {title} {What superconducts in sulfur
  hydrides under pressure and why},\ }\href
  {https://doi.org/10.1103/PhysRevB.91.060511} {\bibfield  {journal} {\bibinfo
  {journal} {Phys. Rev. B}\ }\textbf {\bibinfo {volume} {91}},\ \bibinfo
  {pages} {060511} (\bibinfo {year} {2015})}\BibitemShut {NoStop}%
\bibitem [{\citenamefont {Einaga}\ \emph {et~al.}(2016)\citenamefont {Einaga},
  \citenamefont {Sakata}, \citenamefont {Ishikawa}, \citenamefont {Shimizu},
  \citenamefont {Eremets}, \citenamefont {Drozdov}, \citenamefont {Troyan},
  \citenamefont {Hirao},\ and\ \citenamefont {Ohishi}}]{SH3str2}%
  \BibitemOpen
  \bibfield  {author} {\bibinfo {author} {\bibfnamefont {M.}~\bibnamefont
  {Einaga}}, \bibinfo {author} {\bibfnamefont {M.}~\bibnamefont {Sakata}},
  \bibinfo {author} {\bibfnamefont {T.}~\bibnamefont {Ishikawa}}, \bibinfo
  {author} {\bibfnamefont {K.}~\bibnamefont {Shimizu}}, \bibinfo {author}
  {\bibfnamefont {M.~I.}\ \bibnamefont {Eremets}}, \bibinfo {author}
  {\bibfnamefont {A.~P.}\ \bibnamefont {Drozdov}}, \bibinfo {author}
  {\bibfnamefont {I.~A.}\ \bibnamefont {Troyan}}, \bibinfo {author}
  {\bibfnamefont {N.}~\bibnamefont {Hirao}},\ and\ \bibinfo {author}
  {\bibfnamefont {Y.}~\bibnamefont {Ohishi}},\ }\bibfield  {title} {\bibinfo
  {title} {Crystal structure of the superconducting phase of sulfur hydride},\
  }\href {https://doi.org/10.1038/nphys3760} {\bibfield  {journal} {\bibinfo
  {journal} {Nat. Phys.}\ }\textbf {\bibinfo {volume} {12}},\ \bibinfo {pages}
  {835} (\bibinfo {year} {2016})}\BibitemShut {NoStop}%
\bibitem [{\citenamefont {Li}\ \emph {et~al.}(2016)\citenamefont {Li},
  \citenamefont {Wang}, \citenamefont {Liu}, \citenamefont {Zhang},
  \citenamefont {Hao}, \citenamefont {Pickard}, \citenamefont {Nelson},
  \citenamefont {Needs}, \citenamefont {Li}, \citenamefont {Huang},
  \citenamefont {Errea}, \citenamefont {Calandra}, \citenamefont {Mauri},\ and\
  \citenamefont {Ma}}]{SH3str3}%
  \BibitemOpen
  \bibfield  {author} {\bibinfo {author} {\bibfnamefont {Y.}~\bibnamefont
  {Li}}, \bibinfo {author} {\bibfnamefont {L.}~\bibnamefont {Wang}}, \bibinfo
  {author} {\bibfnamefont {H.}~\bibnamefont {Liu}}, \bibinfo {author}
  {\bibfnamefont {Y.}~\bibnamefont {Zhang}}, \bibinfo {author} {\bibfnamefont
  {J.}~\bibnamefont {Hao}}, \bibinfo {author} {\bibfnamefont {C.~J.}\
  \bibnamefont {Pickard}}, \bibinfo {author} {\bibfnamefont {J.~R.}\
  \bibnamefont {Nelson}}, \bibinfo {author} {\bibfnamefont {R.~J.}\
  \bibnamefont {Needs}}, \bibinfo {author} {\bibfnamefont {W.}~\bibnamefont
  {Li}}, \bibinfo {author} {\bibfnamefont {Y.}~\bibnamefont {Huang}}, \bibinfo
  {author} {\bibfnamefont {I.}~\bibnamefont {Errea}}, \bibinfo {author}
  {\bibfnamefont {M.}~\bibnamefont {Calandra}}, \bibinfo {author}
  {\bibfnamefont {F.}~\bibnamefont {Mauri}},\ and\ \bibinfo {author}
  {\bibfnamefont {Y.}~\bibnamefont {Ma}},\ }\bibfield  {title} {\bibinfo
  {title} {Dissociation products and structures of solid \text{H}$_2$\text{S}
  at strong compression},\ }\href {https://doi.org/10.1103/PhysRevB.93.020103}
  {\bibfield  {journal} {\bibinfo  {journal} {Phys. Rev. B}\ }\textbf {\bibinfo
  {volume} {93}},\ \bibinfo {pages} {020103} (\bibinfo {year}
  {2016})}\BibitemShut {NoStop}%
\bibitem [{\citenamefont {Szab\'o}\ \emph {et~al.}(2001)\citenamefont
  {Szab\'o}, \citenamefont {Samuely}, \citenamefont {Ka\ifmmode \check{c}\else
  \v{c}\fi{}mar\ifmmode~\check{c}\else \v{c}\fi{}\'{\i}k}, \citenamefont
  {Klein}, \citenamefont {Marcus}, \citenamefont {Fruchart}, \citenamefont
  {Miraglia}, \citenamefont {Marcenat},\ and\ \citenamefont
  {Jansen}}]{mgb2-pcs}%
  \BibitemOpen
  \bibfield  {author} {\bibinfo {author} {\bibfnamefont {P.}~\bibnamefont
  {Szab\'o}}, \bibinfo {author} {\bibfnamefont {P.}~\bibnamefont {Samuely}},
  \bibinfo {author} {\bibfnamefont {J.}~\bibnamefont {Ka\ifmmode \check{c}\else
  \v{c}\fi{}mar\ifmmode~\check{c}\else \v{c}\fi{}\'{\i}k}}, \bibinfo {author}
  {\bibfnamefont {T.}~\bibnamefont {Klein}}, \bibinfo {author} {\bibfnamefont
  {J.}~\bibnamefont {Marcus}}, \bibinfo {author} {\bibfnamefont
  {D.}~\bibnamefont {Fruchart}}, \bibinfo {author} {\bibfnamefont
  {S.}~\bibnamefont {Miraglia}}, \bibinfo {author} {\bibfnamefont
  {C.}~\bibnamefont {Marcenat}},\ and\ \bibinfo {author} {\bibfnamefont
  {A.~G.~M.}\ \bibnamefont {Jansen}},\ }\bibfield  {title} {\bibinfo {title}
  {Evidence for two superconducting energy gaps in \text{MgB}$_2$ by
  point-contact spectroscopy},\ }\href
  {https://doi.org/10.1103/PhysRevLett.87.137005} {\bibfield  {journal}
  {\bibinfo  {journal} {Phys. Rev. Lett.}\ }\textbf {\bibinfo {volume} {87}},\
  \bibinfo {pages} {137005} (\bibinfo {year} {2001})}\BibitemShut {NoStop}%
\bibitem [{\citenamefont {Karapetrov}\ \emph {et~al.}(2001)\citenamefont
  {Karapetrov}, \citenamefont {Iavarone}, \citenamefont {Kwok}, \citenamefont
  {Crabtree},\ and\ \citenamefont {Hinks}}]{mgb2-stm}%
  \BibitemOpen
  \bibfield  {author} {\bibinfo {author} {\bibfnamefont {G.}~\bibnamefont
  {Karapetrov}}, \bibinfo {author} {\bibfnamefont {M.}~\bibnamefont
  {Iavarone}}, \bibinfo {author} {\bibfnamefont {W.~K.}\ \bibnamefont {Kwok}},
  \bibinfo {author} {\bibfnamefont {G.~W.}\ \bibnamefont {Crabtree}},\ and\
  \bibinfo {author} {\bibfnamefont {D.~G.}\ \bibnamefont {Hinks}},\ }\bibfield
  {title} {\bibinfo {title} {Scanning tunneling spectroscopy in
  \text{MgB}$_2$},\ }\href {https://doi.org/10.1103/PhysRevLett.86.4374}
  {\bibfield  {journal} {\bibinfo  {journal} {Phys. Rev. Lett.}\ }\textbf
  {\bibinfo {volume} {86}},\ \bibinfo {pages} {4374} (\bibinfo {year}
  {2001})}\BibitemShut {NoStop}%
\bibitem [{\citenamefont {Tsuda}\ \emph {et~al.}(2003)\citenamefont {Tsuda},
  \citenamefont {Yokoya}, \citenamefont {Takano}, \citenamefont {Kito},
  \citenamefont {Matsushita}, \citenamefont {Yin}, \citenamefont {Itoh},
  \citenamefont {Harima},\ and\ \citenamefont {Shin}}]{mgb2-arpes}%
  \BibitemOpen
  \bibfield  {author} {\bibinfo {author} {\bibfnamefont {S.}~\bibnamefont
  {Tsuda}}, \bibinfo {author} {\bibfnamefont {T.}~\bibnamefont {Yokoya}},
  \bibinfo {author} {\bibfnamefont {Y.}~\bibnamefont {Takano}}, \bibinfo
  {author} {\bibfnamefont {H.}~\bibnamefont {Kito}}, \bibinfo {author}
  {\bibfnamefont {A.}~\bibnamefont {Matsushita}}, \bibinfo {author}
  {\bibfnamefont {F.}~\bibnamefont {Yin}}, \bibinfo {author} {\bibfnamefont
  {J.}~\bibnamefont {Itoh}}, \bibinfo {author} {\bibfnamefont {H.}~\bibnamefont
  {Harima}},\ and\ \bibinfo {author} {\bibfnamefont {S.}~\bibnamefont {Shin}},\
  }\bibfield  {title} {\bibinfo {title} {Definitive experimental evidence for
  two-band superconductivity in \text{MgB}$_2$},\ }\href
  {https://doi.org/10.1103/PhysRevLett.91.127001} {\bibfield  {journal}
  {\bibinfo  {journal} {Phys. Rev. Lett.}\ }\textbf {\bibinfo {volume} {91}},\
  \bibinfo {pages} {127001} (\bibinfo {year} {2003})}\BibitemShut {NoStop}%
\bibitem [{\citenamefont {Choi}\ \emph {et~al.}(2002)\citenamefont {Choi},
  \citenamefont {Roundy}, \citenamefont {Sun}, \citenamefont {Cohen},\ and\
  \citenamefont {Louie}}]{mgb2-calnature}%
  \BibitemOpen
  \bibfield  {author} {\bibinfo {author} {\bibfnamefont {H.~J.}\ \bibnamefont
  {Choi}}, \bibinfo {author} {\bibfnamefont {D.}~\bibnamefont {Roundy}},
  \bibinfo {author} {\bibfnamefont {H.}~\bibnamefont {Sun}}, \bibinfo {author}
  {\bibfnamefont {M.~L.}\ \bibnamefont {Cohen}},\ and\ \bibinfo {author}
  {\bibfnamefont {S.~G.}\ \bibnamefont {Louie}},\ }\bibfield  {title} {\bibinfo
  {title} {The origin of the anomalous superconducting properties of
  \text{MgB}$_2$},\ }\href {https://doi.org/10.1038/nature00898} {\bibfield
  {journal} {\bibinfo  {journal} {Nature}\ }\textbf {\bibinfo {volume} {418}},\
  \bibinfo {pages} {758} (\bibinfo {year} {2002})}\BibitemShut {NoStop}%
\bibitem [{\citenamefont {An}\ and\ \citenamefont
  {Pickett}(2001)}]{MgB2_sigma}%
  \BibitemOpen
  \bibfield  {author} {\bibinfo {author} {\bibfnamefont {J.~M.}\ \bibnamefont
  {An}}\ and\ \bibinfo {author} {\bibfnamefont {W.~E.}\ \bibnamefont
  {Pickett}},\ }\bibfield  {title} {\bibinfo {title} {Superconductivity of
  \text{MgB}$_2$: Covalent bonds driven metallic},\ }\href
  {https://doi.org/10.1103/PhysRevLett.86.4366} {\bibfield  {journal} {\bibinfo
   {journal} {Phys. Rev. Lett.}\ }\textbf {\bibinfo {volume} {86}},\ \bibinfo
  {pages} {4366} (\bibinfo {year} {2001})}\BibitemShut {NoStop}%
\bibitem [{\citenamefont {Deringer}\ \emph {et~al.}(2011)\citenamefont
  {Deringer}, \citenamefont {Tchougréeff},\ and\ \citenamefont
  {Dronskowski}}]{coop}%
  \BibitemOpen
  \bibfield  {author} {\bibinfo {author} {\bibfnamefont {V.~L.}\ \bibnamefont
  {Deringer}}, \bibinfo {author} {\bibfnamefont {A.~L.}\ \bibnamefont
  {Tchougréeff}},\ and\ \bibinfo {author} {\bibfnamefont {R.}~\bibnamefont
  {Dronskowski}},\ }\bibfield  {title} {\bibinfo {title} {Crystal orbital
  hamilton population (cohp) analysis as projected from plane-wave basis
  sets},\ }\href {https://doi.org/10.1021/jp202489s} {\bibfield  {journal}
  {\bibinfo  {journal} {J. Phys. Chem. A}\ }\textbf {\bibinfo {volume} {115}},\
  \bibinfo {pages} {5461} (\bibinfo {year} {2011})}\BibitemShut {NoStop}%
\bibitem [{\citenamefont {Maintz}\ \emph {et~al.}(2013)\citenamefont {Maintz},
  \citenamefont {Deringer}, \citenamefont {Tchougréeff},\ and\ \citenamefont
  {Dronskowski}}]{coop2}%
  \BibitemOpen
  \bibfield  {author} {\bibinfo {author} {\bibfnamefont {S.}~\bibnamefont
  {Maintz}}, \bibinfo {author} {\bibfnamefont {V.~L.}\ \bibnamefont
  {Deringer}}, \bibinfo {author} {\bibfnamefont {A.~L.}\ \bibnamefont
  {Tchougréeff}},\ and\ \bibinfo {author} {\bibfnamefont {R.}~\bibnamefont
  {Dronskowski}},\ }\bibfield  {title} {\bibinfo {title} {Analytic projection
  from plane-wave and paw wavefunctions and application to chemical-bonding
  analysis in solids},\ }\href
  {https://doi.org/https://doi.org/10.1002/jcc.23424} {\bibfield  {journal}
  {\bibinfo  {journal} {J. Comput. Chem.}\ }\textbf {\bibinfo {volume} {34}},\
  \bibinfo {pages} {2557} (\bibinfo {year} {2013})}\BibitemShut {NoStop}%
\bibitem [{\citenamefont {Margine}\ and\ \citenamefont
  {Giustino}(2013)}]{ani-ME}%
  \BibitemOpen
  \bibfield  {author} {\bibinfo {author} {\bibfnamefont {E.~R.}\ \bibnamefont
  {Margine}}\ and\ \bibinfo {author} {\bibfnamefont {F.}~\bibnamefont
  {Giustino}},\ }\bibfield  {title} {\bibinfo {title} {Anisotropic
  migdal-eliashberg theory using wannier functions},\ }\href
  {https://doi.org/10.1103/PhysRevB.87.024505} {\bibfield  {journal} {\bibinfo
  {journal} {Phys. Rev. B}\ }\textbf {\bibinfo {volume} {87}},\ \bibinfo
  {pages} {024505} (\bibinfo {year} {2013})}\BibitemShut {NoStop}%
\bibitem [{\citenamefont {Hohenberg}\ and\ \citenamefont {Kohn}(1964)}]{dft1}%
  \BibitemOpen
  \bibfield  {author} {\bibinfo {author} {\bibfnamefont {P.}~\bibnamefont
  {Hohenberg}}\ and\ \bibinfo {author} {\bibfnamefont {W.}~\bibnamefont
  {Kohn}},\ }\bibfield  {title} {\bibinfo {title} {Inhomogeneous electron
  gas},\ }\href {https://doi.org/10.1103/PhysRev.136.B864} {\bibfield
  {journal} {\bibinfo  {journal} {Phys. Rev.}\ }\textbf {\bibinfo {volume}
  {136}},\ \bibinfo {pages} {B864} (\bibinfo {year} {1964})}\BibitemShut
  {NoStop}%
\bibitem [{\citenamefont {Kohn}\ and\ \citenamefont {Sham}(1965)}]{dft2}%
  \BibitemOpen
  \bibfield  {author} {\bibinfo {author} {\bibfnamefont {W.}~\bibnamefont
  {Kohn}}\ and\ \bibinfo {author} {\bibfnamefont {L.~J.}\ \bibnamefont
  {Sham}},\ }\bibfield  {title} {\bibinfo {title} {Self-consistent equations
  including exchange and correlation effects},\ }\href
  {https://doi.org/10.1103/PhysRev.140.A1133} {\bibfield  {journal} {\bibinfo
  {journal} {Phys. Rev.}\ }\textbf {\bibinfo {volume} {140}},\ \bibinfo {pages}
  {A1133} (\bibinfo {year} {1965})}\BibitemShut {NoStop}%
\bibitem [{\citenamefont {Blum}\ \emph {et~al.}(2009)\citenamefont {Blum},
  \citenamefont {Gehrke}, \citenamefont {Hanke}, \citenamefont {Havu},
  \citenamefont {Havu}, \citenamefont {Ren}, \citenamefont {Reuter},\ and\
  \citenamefont {Scheffler}}]{fhi-aims}%
  \BibitemOpen
  \bibfield  {author} {\bibinfo {author} {\bibfnamefont {V.}~\bibnamefont
  {Blum}}, \bibinfo {author} {\bibfnamefont {R.}~\bibnamefont {Gehrke}},
  \bibinfo {author} {\bibfnamefont {F.}~\bibnamefont {Hanke}}, \bibinfo
  {author} {\bibfnamefont {P.}~\bibnamefont {Havu}}, \bibinfo {author}
  {\bibfnamefont {V.}~\bibnamefont {Havu}}, \bibinfo {author} {\bibfnamefont
  {X.}~\bibnamefont {Ren}}, \bibinfo {author} {\bibfnamefont {K.}~\bibnamefont
  {Reuter}},\ and\ \bibinfo {author} {\bibfnamefont {M.}~\bibnamefont
  {Scheffler}},\ }\bibfield  {title} {\bibinfo {title} {Ab initio molecular
  simulations with numeric atom-centered orbitals},\ }\href
  {https://doi.org/https://doi.org/10.1016/j.cpc.2009.06.022} {\bibfield
  {journal} {\bibinfo  {journal} {Comput. Phys. Commun.}\ }\textbf {\bibinfo
  {volume} {180}},\ \bibinfo {pages} {2175} (\bibinfo {year}
  {2009})}\BibitemShut {NoStop}%
\bibitem [{\citenamefont {Perdew}\ \emph {et~al.}(1996)\citenamefont {Perdew},
  \citenamefont {Burke},\ and\ \citenamefont {Ernzerhof}}]{PBE}%
  \BibitemOpen
  \bibfield  {author} {\bibinfo {author} {\bibfnamefont {J.~P.}\ \bibnamefont
  {Perdew}}, \bibinfo {author} {\bibfnamefont {K.}~\bibnamefont {Burke}},\ and\
  \bibinfo {author} {\bibfnamefont {M.}~\bibnamefont {Ernzerhof}},\ }\bibfield
  {title} {\bibinfo {title} {Generalized gradient approximation made simple},\
  }\href {https://doi.org/10.1103/PhysRevLett.77.3865} {\bibfield  {journal}
  {\bibinfo  {journal} {Phys. Rev. Lett.}\ }\textbf {\bibinfo {volume} {77}},\
  \bibinfo {pages} {3865} (\bibinfo {year} {1996})}\BibitemShut {NoStop}%
\bibitem [{\citenamefont {Giustino}(2017)}]{dfptreview}%
  \BibitemOpen
  \bibfield  {author} {\bibinfo {author} {\bibfnamefont {F.}~\bibnamefont
  {Giustino}},\ }\bibfield  {title} {\bibinfo {title} {Electron-phonon
  interactions from first principles},\ }\href
  {https://doi.org/10.1103/RevModPhys.89.015003} {\bibfield  {journal}
  {\bibinfo  {journal} {Rev. Mod. Phys.}\ }\textbf {\bibinfo {volume} {89}},\
  \bibinfo {pages} {015003} (\bibinfo {year} {2017})}\BibitemShut {NoStop}%
\bibitem [{\citenamefont {Baroni}\ \emph {et~al.}(2001)\citenamefont {Baroni},
  \citenamefont {de~Gironcoli}, \citenamefont {Dal~Corso},\ and\ \citenamefont
  {Giannozzi}}]{dfptreview2}%
  \BibitemOpen
  \bibfield  {author} {\bibinfo {author} {\bibfnamefont {S.}~\bibnamefont
  {Baroni}}, \bibinfo {author} {\bibfnamefont {S.}~\bibnamefont
  {de~Gironcoli}}, \bibinfo {author} {\bibfnamefont {A.}~\bibnamefont
  {Dal~Corso}},\ and\ \bibinfo {author} {\bibfnamefont {P.}~\bibnamefont
  {Giannozzi}},\ }\bibfield  {title} {\bibinfo {title} {Phonons and related
  crystal properties from density-functional perturbation theory},\ }\href
  {https://doi.org/10.1103/RevModPhys.73.515} {\bibfield  {journal} {\bibinfo
  {journal} {Rev. Mod. Phys.}\ }\textbf {\bibinfo {volume} {73}},\ \bibinfo
  {pages} {515} (\bibinfo {year} {2001})}\BibitemShut {NoStop}%
\bibitem [{\citenamefont {Giannozzi}\ \emph {et~al.}(2009)\citenamefont
  {Giannozzi}, \citenamefont {Baroni}, \citenamefont {Bonini}, \citenamefont
  {Calandra}, \citenamefont {Car}, \citenamefont {Cavazzoni}, \citenamefont
  {Ceresoli}, \citenamefont {Chiarotti}, \citenamefont {Cococcioni},
  \citenamefont {Dabo}, \citenamefont {Corso}, \citenamefont {de~Gironcoli},
  \citenamefont {Fabris}, \citenamefont {Fratesi}, \citenamefont {Gebauer},
  \citenamefont {Gerstmann}, \citenamefont {Gougoussis}, \citenamefont
  {Kokalj}, \citenamefont {Lazzeri}, \citenamefont {Martin-Samos},
  \citenamefont {Marzari}, \citenamefont {Mauri}, \citenamefont {Mazzarello},
  \citenamefont {Paolini}, \citenamefont {Pasquarello}, \citenamefont
  {Paulatto}, \citenamefont {Sbraccia}, \citenamefont {Scandolo}, \citenamefont
  {Sclauzero}, \citenamefont {Seitsonen}, \citenamefont {Smogunov},
  \citenamefont {Umari},\ and\ \citenamefont {Wentzcovitch}}]{pwscf}%
  \BibitemOpen
  \bibfield  {author} {\bibinfo {author} {\bibfnamefont {P.}~\bibnamefont
  {Giannozzi}}, \bibinfo {author} {\bibfnamefont {S.}~\bibnamefont {Baroni}},
  \bibinfo {author} {\bibfnamefont {N.}~\bibnamefont {Bonini}}, \bibinfo
  {author} {\bibfnamefont {M.}~\bibnamefont {Calandra}}, \bibinfo {author}
  {\bibfnamefont {R.}~\bibnamefont {Car}}, \bibinfo {author} {\bibfnamefont
  {C.}~\bibnamefont {Cavazzoni}}, \bibinfo {author} {\bibfnamefont
  {D.}~\bibnamefont {Ceresoli}}, \bibinfo {author} {\bibfnamefont {G.~L.}\
  \bibnamefont {Chiarotti}}, \bibinfo {author} {\bibfnamefont {M.}~\bibnamefont
  {Cococcioni}}, \bibinfo {author} {\bibfnamefont {I.}~\bibnamefont {Dabo}},
  \bibinfo {author} {\bibfnamefont {A.~D.}\ \bibnamefont {Corso}}, \bibinfo
  {author} {\bibfnamefont {S.}~\bibnamefont {de~Gironcoli}}, \bibinfo {author}
  {\bibfnamefont {S.}~\bibnamefont {Fabris}}, \bibinfo {author} {\bibfnamefont
  {G.}~\bibnamefont {Fratesi}}, \bibinfo {author} {\bibfnamefont
  {R.}~\bibnamefont {Gebauer}}, \bibinfo {author} {\bibfnamefont
  {U.}~\bibnamefont {Gerstmann}}, \bibinfo {author} {\bibfnamefont
  {C.}~\bibnamefont {Gougoussis}}, \bibinfo {author} {\bibfnamefont
  {A.}~\bibnamefont {Kokalj}}, \bibinfo {author} {\bibfnamefont
  {M.}~\bibnamefont {Lazzeri}}, \bibinfo {author} {\bibfnamefont
  {L.}~\bibnamefont {Martin-Samos}}, \bibinfo {author} {\bibfnamefont
  {N.}~\bibnamefont {Marzari}}, \bibinfo {author} {\bibfnamefont
  {F.}~\bibnamefont {Mauri}}, \bibinfo {author} {\bibfnamefont
  {R.}~\bibnamefont {Mazzarello}}, \bibinfo {author} {\bibfnamefont
  {S.}~\bibnamefont {Paolini}}, \bibinfo {author} {\bibfnamefont
  {A.}~\bibnamefont {Pasquarello}}, \bibinfo {author} {\bibfnamefont
  {L.}~\bibnamefont {Paulatto}}, \bibinfo {author} {\bibfnamefont
  {C.}~\bibnamefont {Sbraccia}}, \bibinfo {author} {\bibfnamefont
  {S.}~\bibnamefont {Scandolo}}, \bibinfo {author} {\bibfnamefont
  {G.}~\bibnamefont {Sclauzero}}, \bibinfo {author} {\bibfnamefont {A.~P.}\
  \bibnamefont {Seitsonen}}, \bibinfo {author} {\bibfnamefont {A.}~\bibnamefont
  {Smogunov}}, \bibinfo {author} {\bibfnamefont {P.}~\bibnamefont {Umari}},\
  and\ \bibinfo {author} {\bibfnamefont {R.~M.}\ \bibnamefont {Wentzcovitch}},\
  }\bibfield  {title} {\bibinfo {title} {{QUANTUM} {ESPRESSO}: a modular and
  open-source software project for quantum simulations of materials},\ }\href
  {https://doi.org/10.1088/0953-8984/21/39/395502} {\bibfield  {journal}
  {\bibinfo  {journal} {J. Phys.: Condens. Matter}\ }\textbf {\bibinfo {volume}
  {21}},\ \bibinfo {pages} {395502} (\bibinfo {year} {2009})}\BibitemShut
  {NoStop}%
\bibitem [{\citenamefont {Troullier}\ and\ \citenamefont
  {Martins}(1991)}]{ncpp}%
  \BibitemOpen
  \bibfield  {author} {\bibinfo {author} {\bibfnamefont {N.}~\bibnamefont
  {Troullier}}\ and\ \bibinfo {author} {\bibfnamefont {J.~L.}\ \bibnamefont
  {Martins}},\ }\bibfield  {title} {\bibinfo {title} {Efficient
  pseudopotentials for plane-wave calculations},\ }\href
  {https://doi.org/10.1103/PhysRevB.43.1993} {\bibfield  {journal} {\bibinfo
  {journal} {Phys. Rev. B}\ }\textbf {\bibinfo {volume} {43}},\ \bibinfo
  {pages} {1993} (\bibinfo {year} {1991})}\BibitemShut {NoStop}%
\bibitem [{\citenamefont {Noffsinger}\ \emph {et~al.}(2010)\citenamefont
  {Noffsinger}, \citenamefont {Giustino}, \citenamefont {Malone}, \citenamefont
  {Park}, \citenamefont {Louie},\ and\ \citenamefont {Cohen}}]{epw}%
  \BibitemOpen
  \bibfield  {author} {\bibinfo {author} {\bibfnamefont {J.}~\bibnamefont
  {Noffsinger}}, \bibinfo {author} {\bibfnamefont {F.}~\bibnamefont
  {Giustino}}, \bibinfo {author} {\bibfnamefont {B.~D.}\ \bibnamefont
  {Malone}}, \bibinfo {author} {\bibfnamefont {C.-H.}\ \bibnamefont {Park}},
  \bibinfo {author} {\bibfnamefont {S.~G.}\ \bibnamefont {Louie}},\ and\
  \bibinfo {author} {\bibfnamefont {M.~L.}\ \bibnamefont {Cohen}},\ }\bibfield
  {title} {\bibinfo {title} {\text{EPW}: A program for calculating the
  electron–phonon coupling using maximally localized wannier functions},\
  }\href {https://doi.org/https://doi.org/10.1016/j.cpc.2010.08.027} {\bibfield
   {journal} {\bibinfo  {journal} {Comput. Phys. Commun.}\ }\textbf {\bibinfo
  {volume} {181}},\ \bibinfo {pages} {2140} (\bibinfo {year}
  {2010})}\BibitemShut {NoStop}%
\bibitem [{\citenamefont {Mostofi}\ \emph {et~al.}(2014)\citenamefont
  {Mostofi}, \citenamefont {Yates}, \citenamefont {Pizzi}, \citenamefont {Lee},
  \citenamefont {Souza}, \citenamefont {Vanderbilt},\ and\ \citenamefont
  {Marzari}}]{mlwf}%
  \BibitemOpen
  \bibfield  {author} {\bibinfo {author} {\bibfnamefont {A.~A.}\ \bibnamefont
  {Mostofi}}, \bibinfo {author} {\bibfnamefont {J.~R.}\ \bibnamefont {Yates}},
  \bibinfo {author} {\bibfnamefont {G.}~\bibnamefont {Pizzi}}, \bibinfo
  {author} {\bibfnamefont {Y.-S.}\ \bibnamefont {Lee}}, \bibinfo {author}
  {\bibfnamefont {I.}~\bibnamefont {Souza}}, \bibinfo {author} {\bibfnamefont
  {D.}~\bibnamefont {Vanderbilt}},\ and\ \bibinfo {author} {\bibfnamefont
  {N.}~\bibnamefont {Marzari}},\ }\bibfield  {title} {\bibinfo {title} {An
  updated version of wannier90: A tool for obtaining maximally-localised
  wannier functions},\ }\href
  {https://doi.org/https://doi.org/10.1016/j.cpc.2014.05.003} {\bibfield
  {journal} {\bibinfo  {journal} {Comput. Phys. Commun.}\ }\textbf {\bibinfo
  {volume} {185}},\ \bibinfo {pages} {2309} (\bibinfo {year}
  {2014})}\BibitemShut {NoStop}%
\bibitem [{\citenamefont {Sano}\ \emph {et~al.}(2016)\citenamefont {Sano},
  \citenamefont {Koretsune}, \citenamefont {Tadano}, \citenamefont {Akashi},\
  and\ \citenamefont {Arita}}]{SH3vhs1}%
  \BibitemOpen
  \bibfield  {author} {\bibinfo {author} {\bibfnamefont {W.}~\bibnamefont
  {Sano}}, \bibinfo {author} {\bibfnamefont {T.}~\bibnamefont {Koretsune}},
  \bibinfo {author} {\bibfnamefont {T.}~\bibnamefont {Tadano}}, \bibinfo
  {author} {\bibfnamefont {R.}~\bibnamefont {Akashi}},\ and\ \bibinfo {author}
  {\bibfnamefont {R.}~\bibnamefont {Arita}},\ }\bibfield  {title} {\bibinfo
  {title} {Effect of van hove singularities on high-${T}_{\mathrm{c}}$
  superconductivity in ${\mathrm{h}}_{3}\mathrm{S}$},\ }\href
  {https://doi.org/10.1103/PhysRevB.93.094525} {\bibfield  {journal} {\bibinfo
  {journal} {Phys. Rev. B}\ }\textbf {\bibinfo {volume} {93}},\ \bibinfo
  {pages} {094525} (\bibinfo {year} {2016})}\BibitemShut {NoStop}%
\bibitem [{\citenamefont {Quan}\ and\ \citenamefont {Pickett}(2016)}]{SH3vhs2}%
  \BibitemOpen
  \bibfield  {author} {\bibinfo {author} {\bibfnamefont {Y.}~\bibnamefont
  {Quan}}\ and\ \bibinfo {author} {\bibfnamefont {W.~E.}\ \bibnamefont
  {Pickett}},\ }\bibfield  {title} {\bibinfo {title} {Van hove singularities
  and spectral smearing in high-temperature superconducting
  \text{H}$_3$\text{S}},\ }\href {https://doi.org/10.1103/PhysRevB.93.104526}
  {\bibfield  {journal} {\bibinfo  {journal} {Phys. Rev. B}\ }\textbf {\bibinfo
  {volume} {93}},\ \bibinfo {pages} {104526} (\bibinfo {year}
  {2016})}\BibitemShut {NoStop}%
\bibitem [{\citenamefont {Kong}\ \emph {et~al.}(2001)\citenamefont {Kong},
  \citenamefont {Dolgov}, \citenamefont {Jepsen},\ and\ \citenamefont
  {Andersen}}]{mgb2-ph}%
  \BibitemOpen
  \bibfield  {author} {\bibinfo {author} {\bibfnamefont {Y.}~\bibnamefont
  {Kong}}, \bibinfo {author} {\bibfnamefont {O.~V.}\ \bibnamefont {Dolgov}},
  \bibinfo {author} {\bibfnamefont {O.}~\bibnamefont {Jepsen}},\ and\ \bibinfo
  {author} {\bibfnamefont {O.~K.}\ \bibnamefont {Andersen}},\ }\bibfield
  {title} {\bibinfo {title} {Electron-phonon interaction in the normal and
  superconducting states of \text{MgB}$_{2}$},\ }\href
  {https://doi.org/10.1103/PhysRevB.64.020501} {\bibfield  {journal} {\bibinfo
  {journal} {Phys. Rev. B}\ }\textbf {\bibinfo {volume} {64}},\ \bibinfo
  {pages} {020501} (\bibinfo {year} {2001})}\BibitemShut {NoStop}%
\bibitem [{\citenamefont {Errea}\ \emph {et~al.}(2015)\citenamefont {Errea},
  \citenamefont {Calandra}, \citenamefont {Pickard}, \citenamefont {Nelson},
  \citenamefont {Needs}, \citenamefont {Li}, \citenamefont {Liu}, \citenamefont
  {Zhang}, \citenamefont {Ma},\ and\ \citenamefont {Mauri}}]{SH3anh1}%
  \BibitemOpen
  \bibfield  {author} {\bibinfo {author} {\bibfnamefont {I.}~\bibnamefont
  {Errea}}, \bibinfo {author} {\bibfnamefont {M.}~\bibnamefont {Calandra}},
  \bibinfo {author} {\bibfnamefont {C.~J.}\ \bibnamefont {Pickard}}, \bibinfo
  {author} {\bibfnamefont {J.}~\bibnamefont {Nelson}}, \bibinfo {author}
  {\bibfnamefont {R.~J.}\ \bibnamefont {Needs}}, \bibinfo {author}
  {\bibfnamefont {Y.}~\bibnamefont {Li}}, \bibinfo {author} {\bibfnamefont
  {H.}~\bibnamefont {Liu}}, \bibinfo {author} {\bibfnamefont {Y.}~\bibnamefont
  {Zhang}}, \bibinfo {author} {\bibfnamefont {Y.}~\bibnamefont {Ma}},\ and\
  \bibinfo {author} {\bibfnamefont {F.}~\bibnamefont {Mauri}},\ }\bibfield
  {title} {\bibinfo {title} {High-pressure hydrogen sulfide from first
  principles: A strongly anharmonic phonon-mediated superconductor},\ }\href
  {https://doi.org/10.1103/PhysRevLett.114.157004} {\bibfield  {journal}
  {\bibinfo  {journal} {Phys. Rev. Lett.}\ }\textbf {\bibinfo {volume} {114}},\
  \bibinfo {pages} {157004} (\bibinfo {year} {2015})}\BibitemShut {NoStop}%
\bibitem [{\citenamefont {Bianco}\ \emph {et~al.}(2018)\citenamefont {Bianco},
  \citenamefont {Errea}, \citenamefont {Calandra},\ and\ \citenamefont
  {Mauri}}]{SH3anh2}%
  \BibitemOpen
  \bibfield  {author} {\bibinfo {author} {\bibfnamefont {R.}~\bibnamefont
  {Bianco}}, \bibinfo {author} {\bibfnamefont {I.}~\bibnamefont {Errea}},
  \bibinfo {author} {\bibfnamefont {M.}~\bibnamefont {Calandra}},\ and\
  \bibinfo {author} {\bibfnamefont {F.}~\bibnamefont {Mauri}},\ }\bibfield
  {title} {\bibinfo {title} {High-pressure phase diagram of hydrogen and
  deuterium sulfides from first principles: Structural and vibrational
  properties including quantum and anharmonic effects},\ }\href
  {https://doi.org/10.1103/PhysRevB.97.214101} {\bibfield  {journal} {\bibinfo
  {journal} {Phys. Rev. B}\ }\textbf {\bibinfo {volume} {97}},\ \bibinfo
  {pages} {214101} (\bibinfo {year} {2018})}\BibitemShut {NoStop}%
\end{thebibliography}%

\end{document}